\documentclass[prc,twocolumn,showpacs]{revtex4}
\usepackage{graphicx}
\usepackage{dcolumn}
\usepackage{bm}

\begin{document}

\title{Stellar electron-capture rates in pf-shell nuclei 
from quasiparticle random-phase approximation calculations}

\author{P. Sarriguren}
\affiliation{Instituto de Estructura de la Materia, IEM-CSIC, 
Serrano 123, E-28006 Madrid, Spain}

\email{p.sarriguren@csic.es}

\date{\today}

\begin{abstract}

Electron-capture rates at different density and temperature 
conditions are evaluated for a set of {\it pf}-shell nuclei 
representative of the constituents in presupernova formations.
The nuclear structure part of the problem is described within a
quasiparticle random-phase approximation based on a deformed Skyrme 
Hartree-Fock selfconsistent mean field with pairing correlations 
and residual interactions in particle-hole and particle-particle 
channels. The energy distributions of the Gamow-Teller strength 
are evaluated and compared to benchmark shell-model calculations 
and experimental data extracted from charge-exchange reactions. 
The model dependence of the weak rates are discussed and the 
various sensitivities to both density and temperature are analyzed. 

\end{abstract}


\maketitle

\section{Introduction}

Stars are unique laboratories where all the interactions in nature 
come into play to determine the different stages in the stellar evolution.
This comprises the energy generation, which is mainly caused by fusion 
reactions mediated by strong interactions, as well as the nucleosynthesis 
of elements mediated by nuclear reactions in different stellar scenarios
\cite{b2fh}. Whereas the main sequences of the stellar evolution and the 
majority of the elements with mass number below $A\sim 60$ are produced by 
hydrostatic burning mediated by the strong and electromagnetic interactions,
weak interactions grow in importance in later stages \cite{ffn}, when the 
density ($\rho$) and temperature ($T$) in the core become larger, increasing
the Fermi energy of the degenerate electron gas and then favoring electron 
captures.

In fact, it is a well established feature that weak $\beta$-decay and 
electron-capture (EC) processes are very important mechanisms to understand 
the late stages of the stellar evolution \cite{ffn}, playing a critical role to 
determine both the presupernova stellar structure and the nucleosynthesis 
of heavier nuclei. These processes are dominated by Gamow-Teller (GT) 
transitions and therefore, the GT properties of {\it pf}-shell nuclei are of 
special importance because they are the main constituents of the stellar 
core in presupernovae formations \cite{aufderheide-94} leading to
core-collapse (type II) or thermonuclear (type Ia) supernovae.

While the scenarios for type Ia supernovae are thought to be binary 
systems with a white dwarf accreting material from a companion star,
type II supernovae are the final result of the gravitational collapse of 
the core of a massive star that takes place when the nuclear fuel exhausts.
Then, the core becomes unstable and when the mass exceeds the Chandrasekhar 
mass, the electron degeneracy pressure is not able to prevent the 
gravitational collapse. In the initial stages, electrons are captured by 
nuclei in the iron-nickel mass region, thus reducing the electron-to-baryon 
fraction ($Y_e$) of the presupernova star and correspondingly the degeneracy 
pressure. At the same time, the neutrinos produced at presupernova densities
leave the star reducing the energy and cooling the star. Both effects act in 
the same direction accelerating the collapse. 
With increasing neutronization of the core material, the $Q_\beta$ energies 
increase and $\beta$-decays become more important competing with ECs.
EC processes are therefore essential ingredients to follow the complex 
dynamics of core-collapse supernovae and reliable estimates of these rates 
certainly contribute to a better understanding of the explosion mechanism.

An accurate understanding of most astrophysical processes requires necessarily 
information from nuclear physics, which provides the input to deal with network 
calculations and astrophysical simulations \cite{langanke-03}.
Due to the extreme conditions of $\rho$ and  $T$ that hold in stellar scenarios, 
most of the nuclear properties cannot be measured directly. Therefore, the GT 
strength distributions must be estimated in many cases by model calculations. 
Obviously, nuclear physics uncertainties will finally affect the reliability 
of the description of those astrophysical processes.

The first extensive calculations of stellar weak rates as functions of 
relevant ranges of $\rho$ and $T$ were done in Ref. \cite{ffn}. It was 
assumed  that the whole GT strength resides in a single resonance whose 
energy relative to the daughter ground state is parametrized 
phenomenologically, taking the total GT strength from the single-particle 
model. 

In the last decades, GT$^+$ strength distributions on nuclei in the mass 
region $A\sim 60$ have been studied experimentally via $(n,p)$, or equivalent
higher resolution charge-exchange reactions such as $(d,^2{\rm He})$ and 
$(t,^3{\rm He})$, at forward angles 
\cite{alford-91,yako-09,rakers-04,alford-93,baumer-03,ronnqvist-93,
vetterli-87,elkateb-94,hagemann-04,cole-06,williams-95,anantaraman-08,
popescu-07,grewe-08,hitt-09,fujita-11}.
Charge-exchange reactions are the most efficient way to extract the GT$^+$ 
strength in stable nuclei \cite{osterfeld-92}. For incident energies above 
100 MeV, the isovector spin-flip component of the effective interaction is 
dominant and the cross sections is mainly originated from spin-isospin 
transitions. At forward angles the momentum transfer is small and the 
reaction cross section is dominated by the GT operator with 
$\Delta T=1, \Delta L=0, \Delta J^\pi=1^+$. The cross section, 
extrapolated to zero momentum transfer, is proportional to the 
$\beta$-decay strength between the same states. Charge-exchange 
reactions at small momentum transfer are therefore used to study 
GT strength distributions when $\beta$-decay is not energetically 
possible.

The data show that the total GT$^+$ strength is strongly quenched and 
fragmented over many final states, as compared to the independent-particle 
model. Then, improvements in the weak rates have been focused 
on the description of the nuclear structure aspect of the problem. 
Different approaches to describe the nuclear structure involved in the 
stellar weak decay rates can be found in the literature. 
They are basically divided into shell model (SM), either SM Montecarlo 
\cite{koonin-97} or large scale SM diagonalizations 
\cite{langanke-00,langanke-01,suzuki-09} and proton-neutron 
quasiparticle random-phase approximation (QRPA) 
\cite{moeller-97,nabi-99,nabi-09,paar-04,sarri-09-11,fantina-12} 
categories. 
Although QRPA calculations cannot reach the detailed spectroscopy 
achieved from present state-of-the-art SM calculations,
the global performance of QRPA is quite satisfactory. Moreover, 
one clear advantage of the QRPA method is that it can be extended 
to heavier nuclei, which are beyond the present capability of full 
SM calculations, without increasing the complexity of the calculation. 

Very recently, a systematic evaluation of the ability to reproduce the 
measured GT strength distributions of various theoretical models based 
on SM and QRPA was done in Ref. \cite{cole-12}, where EC rates were 
derived from those models at relevant $\rho$ and $T$.
While several sets of SM calculations using different effective
interactions were compared, namely KB3G \cite{poves-01} and GXPF1a
\cite{honma-02}, in the case of QRPA only the formalism developed in
\cite{krumlinde-84} using deformations and masses obtained from the 
finite range droplet model \cite{moeller-95} was considered in 
Ref. \cite{cole-12}. In what follows we use the term QRPA-M\"oller to 
refer to these QRPA results obtained from the above mentioned formalism.

Given the sensitivity of the weak rates to the nuclear structure 
through the GT strength distributions, it is worth extending the
study in Ref. \cite{cole-12} by considering alternative calculations
within the QRPA formalism.
Indeed, the QRPA method with separable GT interactions was first 
proposed and applied in Ref. \cite{halbleib-67}, on a spherical harmonic 
oscillator basis, and then it was extended to deformed nuclei 
\cite{krumlinde-84} using deformed phenomenological single-particle basis. 
Further refinements to the QRPA formalism were introduced along the years 
\cite{muto-89,frisk-95,sarri-98,sarri-99,sarri-01,sarri-01-odd,borzov-06}.

In this work, we study the dependence of the EC rates on both $\rho$ and 
$T$ with GT strength distributions calculated within a QRPA approach 
based on a selfconsistent deformed Hartree-Fock (HF) mean field with 
Skyrme interactions including pairing correlations and residual
separable forces in both particle-hole ($ph$) and particle-particle 
($pp$) channels. We compare our calculations with the benchmark
calculations in Ref. \cite{cole-12}.
This formalism represents an improvement over the QRPA-M\"oller approach 
in several aspects. First, instead of a phenomenological approach 
(single-particle models based on Nilsson, Woods-Saxon, or folded 
Yukawa models), the deformed mean field is now obtained selfconsistently 
and there is no need to introduce deformation parameters as input, and 
secondly, a separable residual GT interaction in the $pp$ channel is 
included. We should also mention the inclusion of an effective quenching 
factor in our calculations that was not included in QRPA-M\"oller.
The present nuclear model has been tested successfully reproducing very
reasonably the experimental information available on both bulk and
decay properties of medium-mass nuclei 
\cite{sarri-03,poirier-04,nacher-04,sarri-05-wp,sarri-09-prc,sarri-10}.

The paper is organized as follows. In Section \ref{form} the weak decay 
rates are introduced as functions of $\rho$ and $T$ and their nuclear 
structure and phase space components are described. 
Section \ref{results} contains the results obtained for the GT strength 
distributions and for the EC rates in some selected {\it pf}-shell nuclei 
that correspond to those used in Ref. \cite{cole-12}. 
Section \ref{conclusions} contains the conclusions of this work.

\section{Theoretical Formalism}
\label{form}

There are several distinctions between terrestrial and stellar decay
rates caused by the effect of high  $\rho$ and $T$. One effect of
$T$ is directly related to the thermal population of excited states in
the decaying nucleus, accompanied by the corresponding depopulation of
the ground states. The weak-decay rates of excited states can be
significantly different from those of the ground state and a
case by case consideration is needed. Another effect related to the
high $\rho$ and $T$ comes from the fact that atoms in these scenarios
are completely ionized and consequently electrons are no longer bound
to the nuclei, but forming a degenerate plasma obeying a Fermi-Dirac
distribution. This opens the possibility for continuum EC, in contrast 
to the orbital EC caused by bound electrons in the atom under 
terrestrial conditions.
These effects make weak interaction rates in the stellar interior
sensitive functions of $T$ and $\rho$.

\subsection{ Weak decay rates}  
\label{wdr}

Assuming thermal equilibrium, the probability of occupation of the 
excited states in the parent nucleus follows a Boltzmann distribution.
The decay rate of the parent nucleus is given by

\begin{equation}
\lambda = \sum_i \lambda_i\, \frac{2J_i+1}{G} e^{-E_i/(k_BT)} \, ,
\label{population}
\end{equation}
where $G=\sum_i \left( 2J_i+1 \right) e^{-E_i/(k_BT)}$ is the partition
function and $J_i(E_i)$ is the angular momentum (excitation energy) of 
the parent nucleus state $i$.

In principle, the sum extends over all populated states in the parent
nucleus up to the proton separation energy. However, because of the
range of temperatures considered in this work ($T=1-10$ GK), only a 
few low-lying excited states are expected to contribute 
in the decay of even-even nuclei. In fact, the lowest excited states
in the even-even nuclei considered are collective $2^+$ states located 
typically above 1 MeV from the ground state. The lowest of these $2^+$ 
states corresponds to the nucleus $^{56}$Fe and appears at 
$E_{2^+}=0.847$ MeV  \cite{ensdf}. Hence, their contributions to the 
rates can be neglected at these temperatures. 

In the case of odd-$A$ nuclei the situation is more involved because 
excited states of single-particle nature appear at low excitation energies.
This is particularly the case of $^{45}$Sc, where one finds up to six
excited states below 1 MeV \cite{ensdf}. These states will become
populated as $T$ raises and may contribute to the weak rates. A case by 
case analysis is mandatory and work in this line is in progress.

The decay rate corresponding to the parent state $i$ is given by

\begin{equation}
\lambda _i = \sum_f \lambda_{if} =  \frac{\ln 2}{D}  \sum_f 
B_{if}\Phi_{if} (\rho,T)\, ,
\end{equation}
where the sum extends over all the states in the final nucleus reached 
in the decay process and $D=6146$ s. This expression is decomposed into 
a phase space factor $\Phi_{if}$, which is a function of 
$\rho$ and $T$ and a nuclear structure part $B_{if}$ that contains the 
transition probabilities for allowed Fermi and GT transitions,

\begin{equation}
B_{if}=B_{if}(GT)+ B_{if}(F)\, .
\end{equation}
In this work we only consider the dominant GT transitions. Fermi 
transitions have a simple expression assuming isospin symmetry and 
are only important for $\beta^+$ decay of neutron-deficient light 
nuclei with  $Z> N$. The theoretical description of both 
$B_{if}$ and $\Phi_{if}$ are explained in the next subsections.

\subsection{Phase Space Factors}

The nuclei under study in this work correspond to stable {\it pf}-shell 
nuclei. $\beta^+$ decays from their ground states are then energetically 
forbidden. However, as $T$ raises, thermal population of excited states 
in the parent nucleus may induce $\beta ^+$-decays if $E_i$ exceeds the 
$Q_{\beta ^+}$ energy. These decays that are almost independent of $\rho$ 
and $T$, might compete with ECs in cases where $T$ is high enough to 
populate significantly excited states in the parent nucleus. 
In our case the decays would involve excited states beyond 1.5-2 MeV 
in the most favored cases, but they are not sufficiently populated 
even at the higher $T$ considered in this work ($k_BT=0.862$ MeV at 
$T_9=10$ with $T_9=10^9$K) Contributions to the weak rates from positron 
decays are therefore neglected in this work.

In the astrophysical scenarios of our study, nuclei are fully ionized 
and continuum EC from the degenerate electron plasma are possible. 
The phase space factor for EC is given by

\begin{eqnarray}
\Phi^{EC}_{if}&=&\int_{\omega_\ell}^{\infty} \omega p (Q_{if}+\omega)^2
F(Z,\omega) \nonumber \\
&& \times S_{e}(\omega) \left[ 1-S_{\nu}(Q_{if}+\omega)\right] d\omega \, .
\label{phiec}
\end{eqnarray}
In this expression $\omega$ is the total energy of the electron in 
$m_ec^2$ units, $p=\sqrt{\omega ^2 -1}$ is the momentum,
and $Q_{if}$ is the total energy available in $m_e c^2$ units

\begin{equation}
Q_{if}=\frac{1}{m_ec^2}\left( Q_{EC} - m_e c^2 +E_i-E_f \right) \, ,
\label{qif}
\end{equation}
with 
\begin{equation}
Q_{EC} = Q_{\beta^+} + 2m_e c^2 = \left( M_p-M_d+m_e\right) c^2 \, ,
\label{qec}
\end{equation}
written in terms of the nuclear masses of parent ($M_p$)
and daughter ($M_d$) nuclei and their excitation energies $E_i$ and
$E_f$, respectively. 

$F(Z,\omega)$ is the Fermi function
that takes into account the distortion of the electron wave
function due to the Coulomb interaction.

\begin{equation}
F(Z,\omega ) = 2(1+\gamma) (2pR)^{-2(1-\gamma)} e^{\pi y}
\frac{|\Gamma (\gamma+iy)|^2}{[\Gamma (2\gamma+1)]^2}\, ,
\end{equation}
where $\gamma=\sqrt{1-(\alpha Z)^2}$ ; $y=\alpha Z\omega /p$ ; 
$\alpha$ is the fine structure constant and $R$ the nuclear radius. 
The lower integration limit in Eq. (\ref{phiec}) is given by
${\omega_\ell}=1$ if $Q_{if}> -1$, or ${\omega_\ell}=|Q_{if}|$ if 
$Q_{if}< -1$.

$S_e$ and $S_\nu$, are the electron and neutrino distribution functions, 
respectively. Its presence inhibits or enhances the phase space available. 
In the stellar scenarios considered here the commonly accepted assumption 
is that $S_\nu=0$, because neutrinos and antineutrinos at these densities
can escape freely from the interior of the star. The electron distribution 
is described as a Fermi-Dirac distribution

\begin{equation}
S_{e}=\frac{1}{\exp \left[ \left(\omega -\mu_e\right)/(k_BT)\right] +1} \, .
\end{equation}
The chemical potential $\mu_e$ as a function of $\rho$ and $T$ is 
determined from the expression

\begin{equation}
\rho Y_e = \frac{1}{\pi^2 N_A}\left( \frac{m_e c}{\hbar}\right) ^3 
\int_0^{\infty} (S_e - S_{e^+}) p^2 dp \, ,
\end{equation}
in (mol/cm$^3$) units. $\rho$ is the baryon density (g/cm$^3$),
$Y_e$ is the electron-to-baryon ratio (mol/g), and $N_A$ is Avogadro's
number (mol$^{-1}$).

The phase space factor for EC in Eq. (\ref{phiec}) is therefore a 
sensitive function of both $\rho$ and $T$, through the electron 
distribution $S_e$. 

\subsection{Nuclear Structure}

The nuclear structure part of the problem is described within the
QRPA formalism. Various approaches have been developed in the past
to describe the spin-isospin nuclear excitations in QRPA 
\cite{paar-04,krumlinde-84,muto-89,frisk-95,sarri-98,sarri-99,
sarri-01,sarri-01-odd,borzov-06}.
In this subsection we show briefly the theoretical framework used in
this work to describe the nuclear part of the decay rates.
More details of the
formalism can be found in Refs. \cite{sarri-98,sarri-01,sarri-01-odd}.

The method starts with a self-consistent deformed Hartree-Fock mean 
field formalism obtained with Skyrme interactions, including
pairing correlations. The single-particle energies, wave functions,
and occupation probabilities are generated from this mean field.
In this work we have chosen the Skyrme force SLy4 \cite{sly4} as a
representative of the Skyrme forces. It is one of the most successful 
Skyrme forces and has been extensively studied in the last years.
We also consider for comparison the results obtained with the force 
SG2 \cite{giai} that has been successfully tested against spin-isospin 
excitations in spherical and deformed nuclei.
 
The solution of the HF equation is found by using the formalism 
developed in Ref. \cite{vautherin}, assuming time reversal and axial 
symmetry. The single-particle wave functions are expanded in terms 
of the eigenstates of an axially symmetric harmonic oscillator in 
cylindrical coordinates, using twelve major shells. The method also 
includes pairing between like nucleons in BCS approximation with 
fixed gap parameters for protons and neutrons, which are determined
phenomenologically from the odd-even mass differences involving
the experimental binding energies \cite{ensdf}. 

Potential energy curves are analyzed as a function of
the quadrupole deformation. For that purpose, constrained 
HF calculations are performed with a quadratic constraint 
\cite{flocard-73}. The HF energy is minimized under the constraint 
of keeping fixed the nuclear deformation. Calculations for GT 
strengths are performed subsequently for the various equilibrium
shapes of each nucleus, that is, for the solutions, in general
deformed, for which minima are obtained in the energy curves. 

To describe GT transitions, a spin-isospin residual interaction is
added to the Skyrme mean field and treated in a deformed proton-neutron 
QRPA. Taking into account the phenomenological nature of the Skyrme
interactions, which are not fitted to spin-isospin observables, it is
reasonable to complement the Skyrme interaction with some extra 
phenomenological parameters sensitive to those nuclear properties.

This interaction contains a $ph$ and a $pp$ part. The interaction in 
the $ph$ channel is mainly responsible for the position and structure 
of the GT resonance and it can be derived consistently from 
the same Skyrme interaction used to generate the mean field, through 
the second derivatives of the energy density functional with respect 
to the one-body densities. The $ph$ residual interaction is finally 
expressed in a separable form by averaging the resulting contact 
force over the nuclear volume \cite{sarri-98}. 
By taking separable GT forces, the energy eigenvalue problem 
reduces to find the roots of an algebraic equation. 

The $pp$ part is a neutron-proton pairing force in the $J^\pi=1^+$ 
coupling channel, which is also introduced as a separable force 
\cite{muto-89,sarri-01}. Its strength is usually fitted to 
reproduce globally the experimental half-lives. 
Various attempts have been done in the past to fix this strength 
\cite{homma-96}, arriving to expressions that depend on the model used 
to describe the mean field, Nilsson model in the above reference. 
In previous works \cite{sarri-98,sarri-99,sarri-01,sarri-01-odd,sarri-03}
we have studied the sensitivity of the GT strength distributions to 
the various ingredients contributing to the deformed QRPA 
calculations, namely to the nucleon-nucleon effective force, to 
pairing correlations, and to residual interactions. We found different 
sensitivities to them. In this work, all of these ingredients have been 
fixed to the most reasonable choices found previously. 
In particular we use the coupling strengths 
$\chi ^{ph}_{GT}=0.10$ MeV and $\kappa ^{pp}_{GT} = 0.05$ MeV.
An optimum set of coupling strengths $(\chi ^{ph}_{GT}, \kappa ^{pp}_{GT})$ 
could be chosen following a case by case fitting procedure and we will 
get different answers depending on the nucleus, shape, and Skyrme force. 
However, since the purpose here is to test the ability of QRPA to account 
for the GT strength distributions in the iron-nickel mass region with as 
few free parameters as possible, we have chosen to use the same coupling 
strengths for all the nuclei considered in this work. 

The GT strength for a transition from an initial state $i$ to a
final state $f$ is given by

\begin{equation}
B_{if}(GT^{\pm} )= \frac{1}{2J_i+1} \left( \frac{g_A}{g_V} 
\right)_{\rm eff} ^2 \langle f || \sum_j^A \sigma_j t^{\pm}_j 
|| i \rangle ^2 \, ,
\end{equation}
$(g_A/g_V)_{\rm eff} = 0.7 (g_A/g_V)_{\rm bare}$  is the
effective ratio of axial and vector coupling factors that takes into
account in an effective manner the observed quenching of the GT strength. 

In even-even nuclei, the GT strength corresponding to the transition 
$J_iK_i \rightarrow J_fK_f $ $(0^+0 \rightarrow 1^+K$) in the laboratory
system, is expressed in terms of the intrinsic amplitudes connecting the
QRPA ground state  $\left| \phi_0\right>$ with the one-phonon excited 
states $\left| \phi_K\right>$

\begin{eqnarray}
B_{if}(GT^{\pm} )=\left( \frac{g_A}{g_V} \right)_{\rm eff} ^2 
&& \left[ \delta_{K,0}
\left< \phi_{K} \left|  \sigma_0t^\pm  \right| \phi_0\right> ^2 \right.\\
&&+2\left. \delta_{K,1}
\left< \phi_{K} \left|  \sigma_1t^\pm  \right| \phi_0\right> ^2 \right] 
\, .
\label{streven}
\end{eqnarray}
To obtain this expression, the initial and final states in the
laboratory frame have been expressed in terms of the intrinsic states
using the Bohr-Mottelson factorization \cite{bm}.

When the parent nucleus has an odd nucleon, the
ground state can be expressed as a one quasi-particle state in which the 
odd nucleon occupies the single-particle orbit of lowest energy. 
Then two types of transitions are possible. One type is due to phonon
excitations in which the odd nucleon acts only as a spectator. In the 
intrinsic frame, the transition amplitudes are in this case basically 
the same as in the even-even case  but with the blocked spectator 
excluded from the calculation.
The other type of transitions are those involving the odd nucleon state,
which are treated by taking into 
account phonon correlations in the quasiparticle transitions in first 
order perturbation \cite{krumlinde-84,muto-89,sarri-01}.

\section{Results}
\label{results}

\begin{figure}
\includegraphics[width=0.39\textwidth]{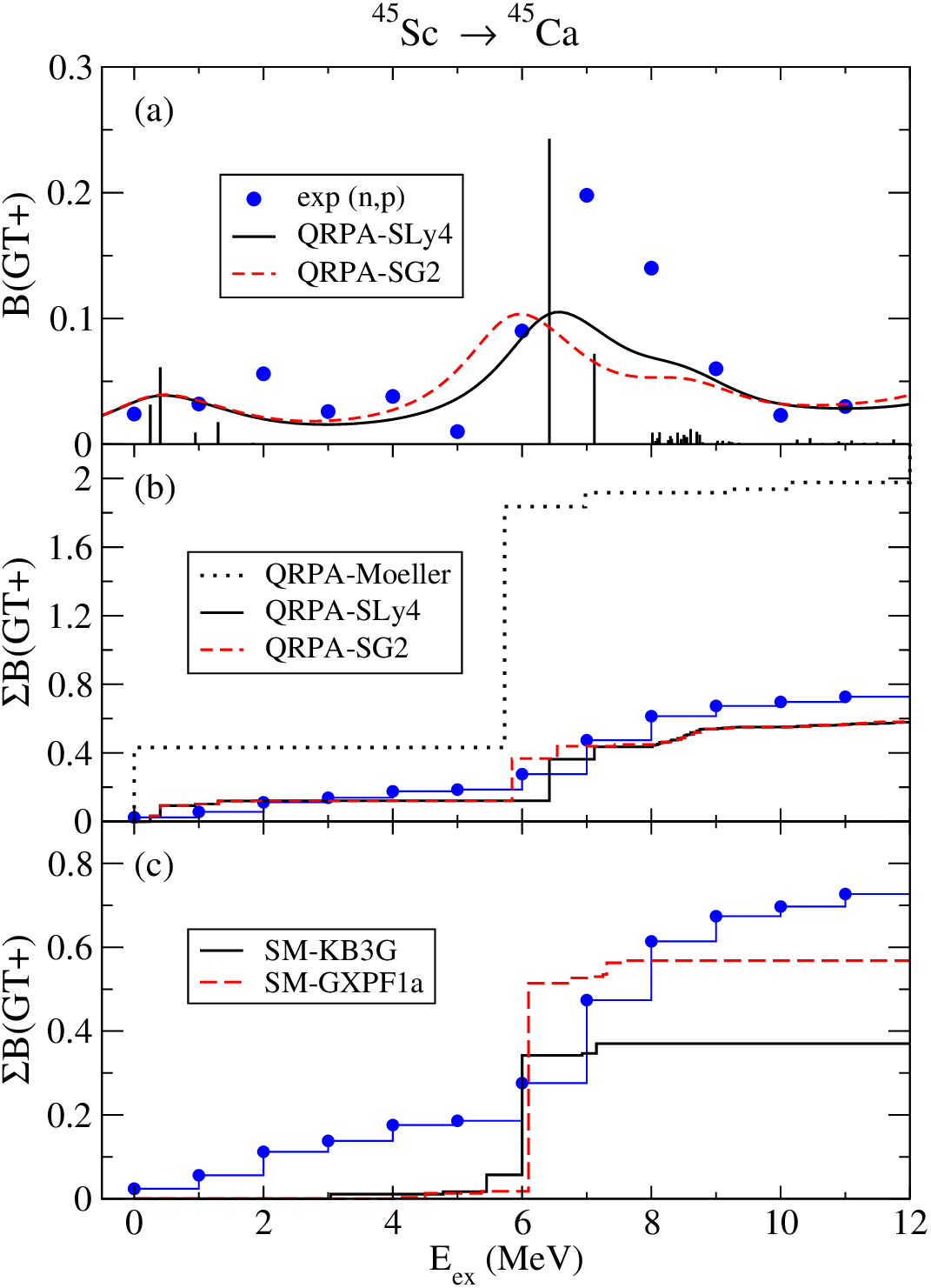}
\caption{(Color online) Gamow-Teller strength distribution B(GT$^+$) for 
the transition $^{45}$Sc to $^{45}$Ca plotted versus the excitation energy 
of the daughter nucleus. (a) Experimental data are compared to our QRPA 
results with SLy4 Skyrme force (individual transitions) as well as with 
folded distributions from SLy4 and SG2 forces. (b) Experimental accumulated 
B(GT$^+$) strength compared to QRPA results from M\"oller \cite{krumlinde-84} 
as given in Ref. \cite{cole-12} and from our calculations with SLy4 and SG2 
interactions. (c) Experimental accumulated B(GT$^+$) strength compared to 
shell-model results with KB3G and GXPF1a interactions.
Data extracted from $(n,p)$ reactions are from Ref. \cite{alford-91}.} 
\label{bgt_sc45}
\end{figure}

\begin{figure}
\includegraphics[width=0.39\textwidth]{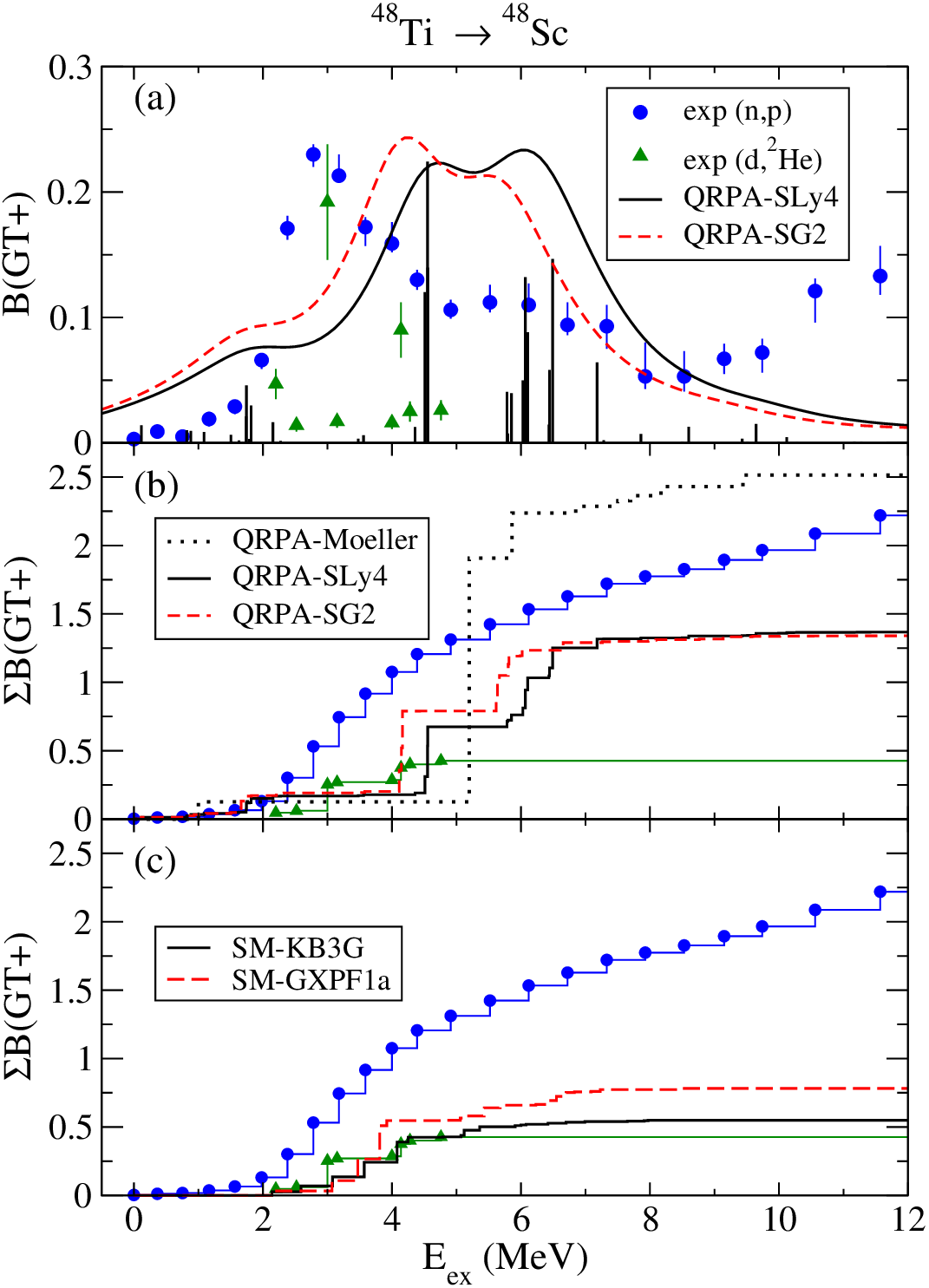}
\caption{(Color online) Same as in Fig. \ref{bgt_sc45}, but for the
$^{48}$Ti to $^{48}$Sc transition.
Data are from $(n,p)$ \cite{yako-09} and $(d,^2{\rm He})$ \cite{rakers-04} 
reactions.} 
\label{bgt_ti48}
\end{figure}

\begin{figure}
\includegraphics[width=0.39\textwidth]{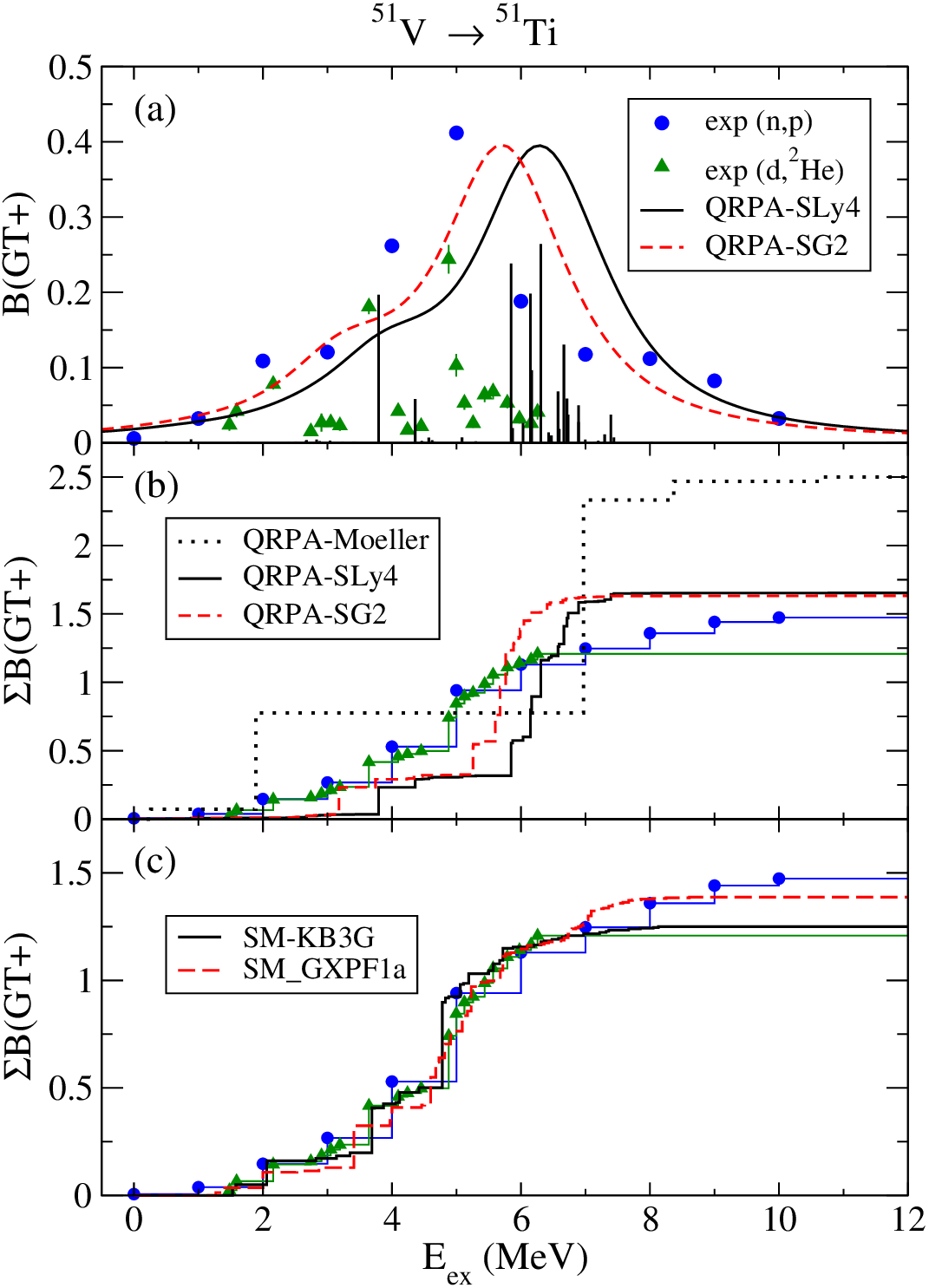}
\caption{(Color online) Same as in Fig. \ref{bgt_sc45}, but for the
$^{51}$V to $^{51}$Ti transition.
Data are from $(n,p)$ \cite{alford-93} and $(d,^2{\rm He})$ \cite{baumer-03} 
reactions.} 
\label{bgt_v51}
\end{figure}

\begin{figure}
\includegraphics[width=0.39\textwidth]{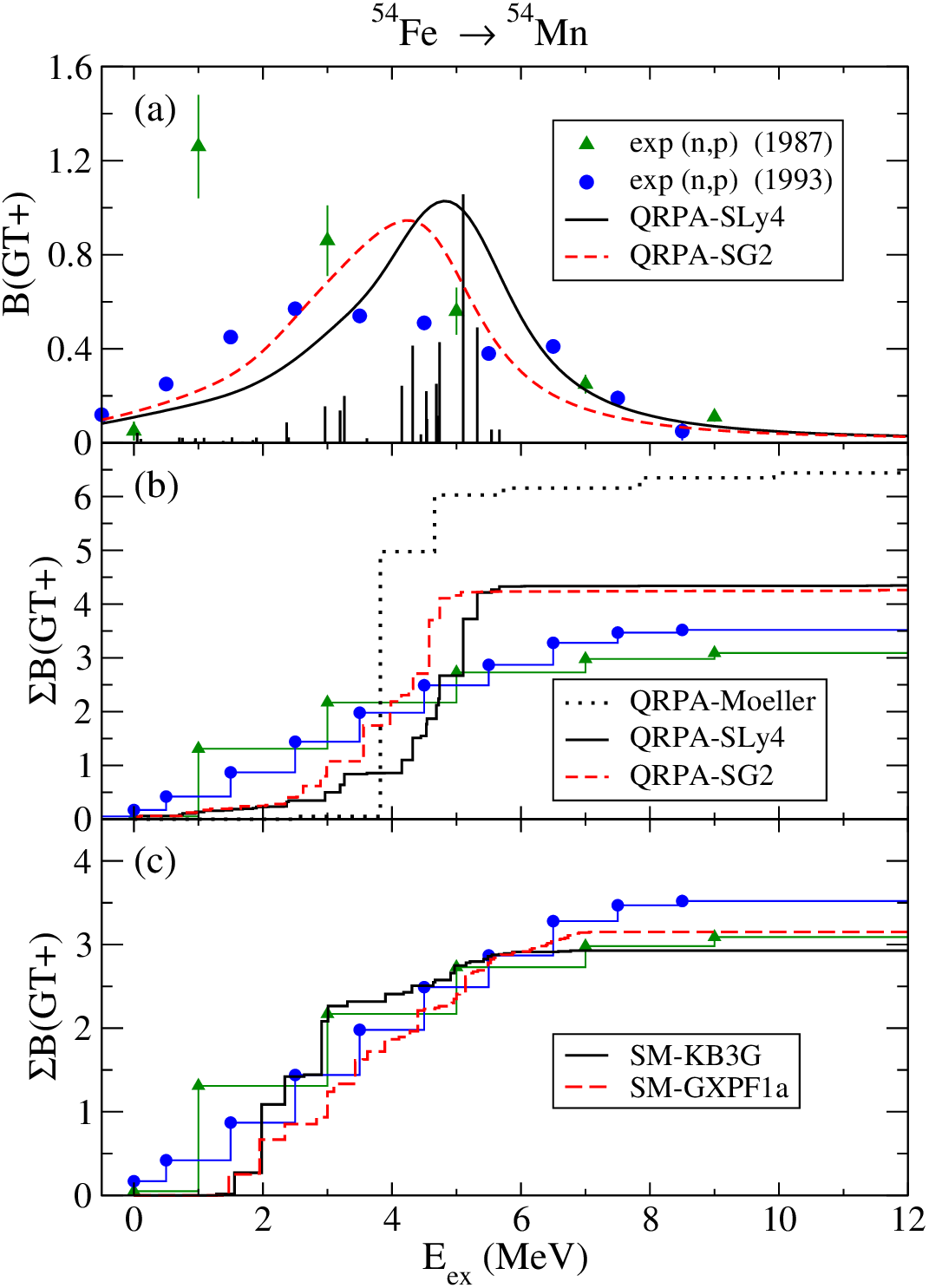}
\caption{(Color online) Same as in Fig. \ref{bgt_sc45}, but for the
$^{54}$Fe to $^{54}$Mn transition.
Data are from $(n,p)$ reactions from Refs. \cite{ronnqvist-93,vetterli-87}}.
\label{bgt_fe54}
\end{figure}

\begin{figure}
\includegraphics[width=0.39\textwidth]{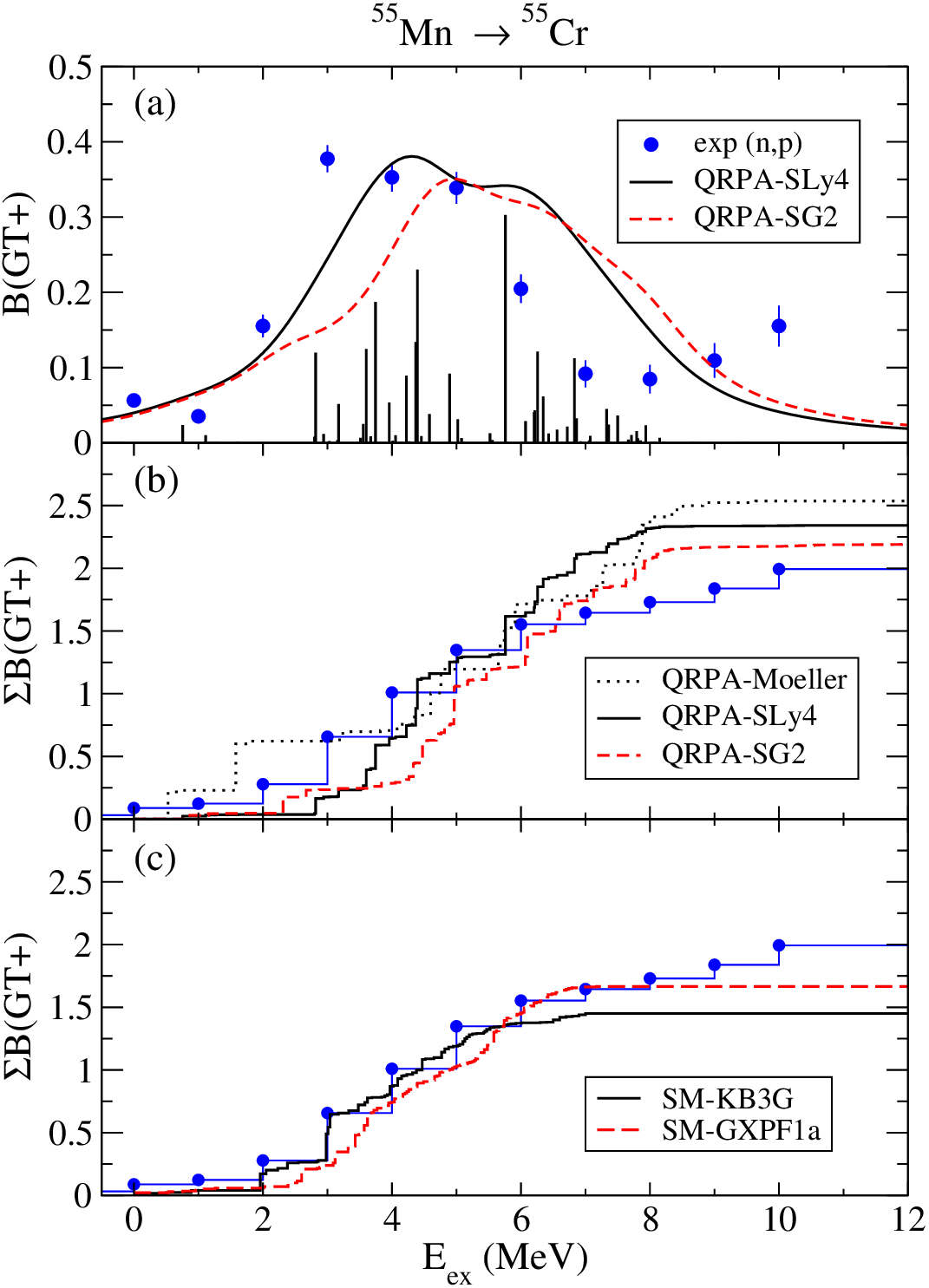}
\caption{(Color online) Same as in Fig. \ref{bgt_sc45}, but for the
$^{55}$Mn to $^{55}$Cr transition.
Data are from $(n,p)$ reactions \cite{elkateb-94}.} 
\label{bgt_mn55}
\end{figure}

\begin{figure}
\includegraphics[width=0.39\textwidth]{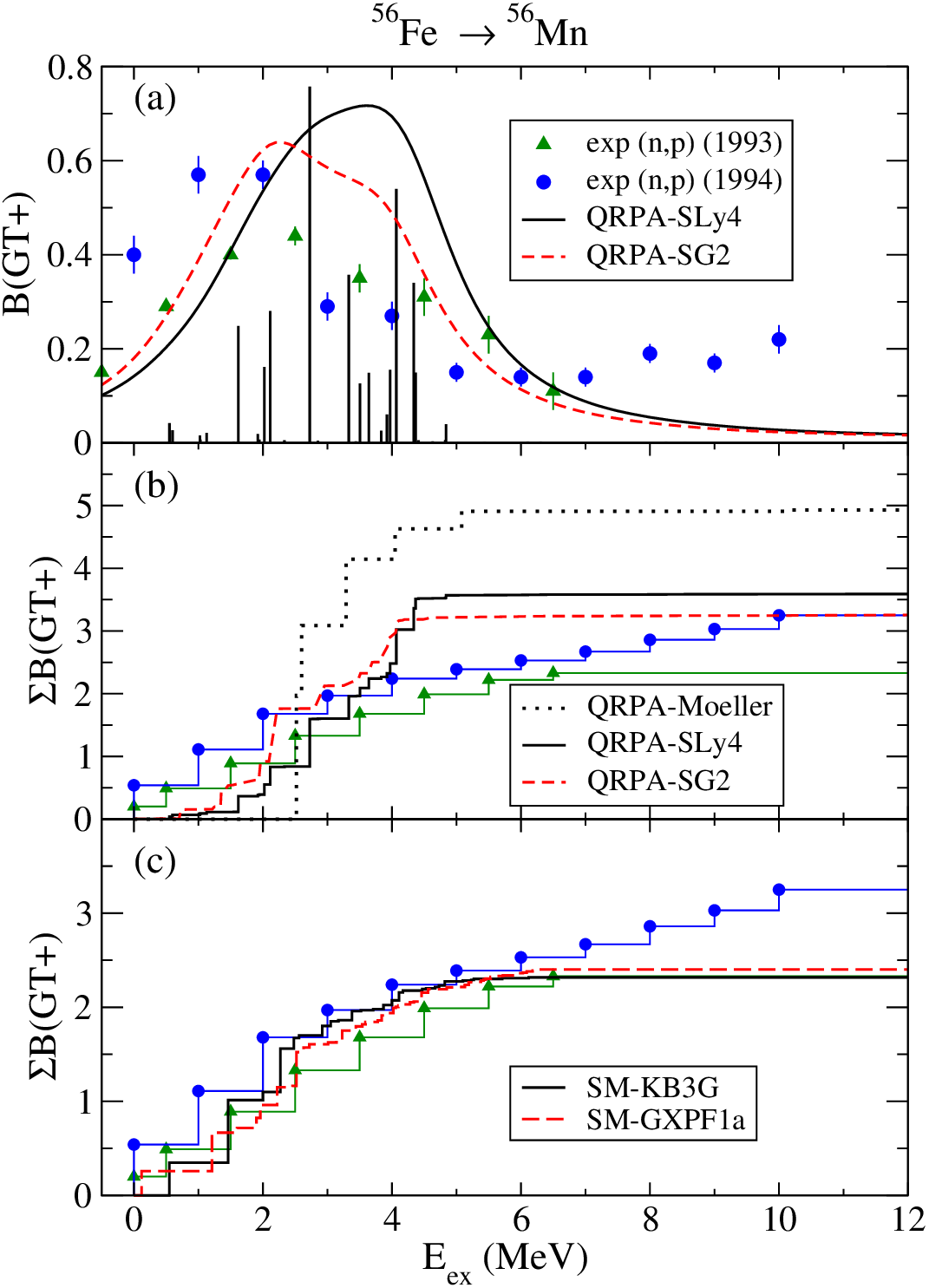}
\caption{(Color online) Same as in Fig. \ref{bgt_sc45}, but for the
$^{56}$Fe to $^{56}$Mn transition.
Data are from $(n,p)$ reactions from Refs. \cite{ronnqvist-93,elkateb-94}}.
\label{bgt_fe56}
\end{figure}

\begin{figure}
\includegraphics[width=0.39\textwidth]{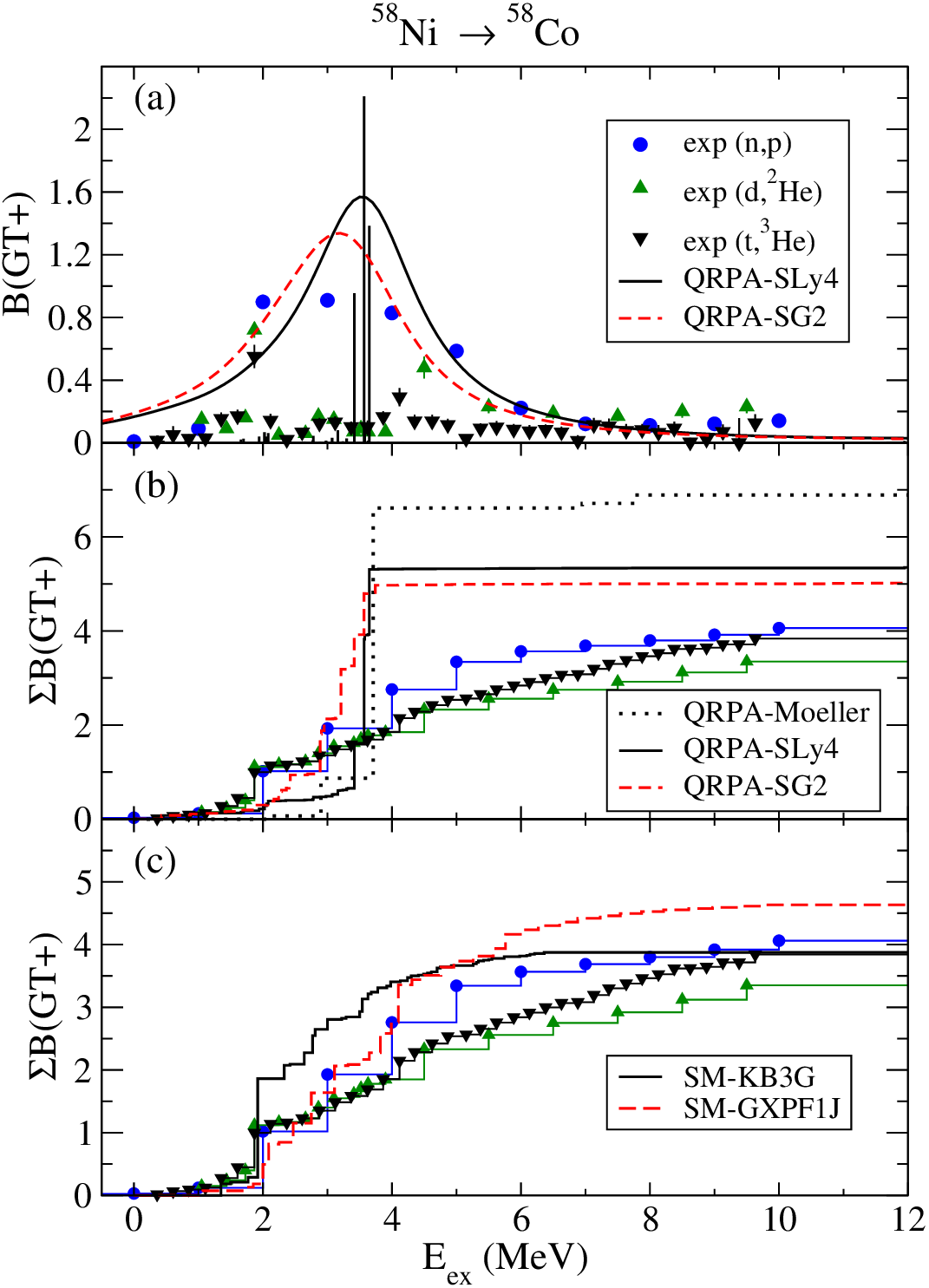}
\caption{(Color online) Same as in Fig. \ref{bgt_sc45}, but for the
$^{58}$Ni to $^{58}$Co transition.
Data are from $(n,p)$ \cite{elkateb-94}, $(d,^2{\rm He})$ \cite{hagemann-04}, 
and $(t,^3{\rm He})$ \cite{cole-06} reactions.} 
\label{bgt_ni58}
\end{figure}

\begin{figure}
\includegraphics[width=0.39\textwidth]{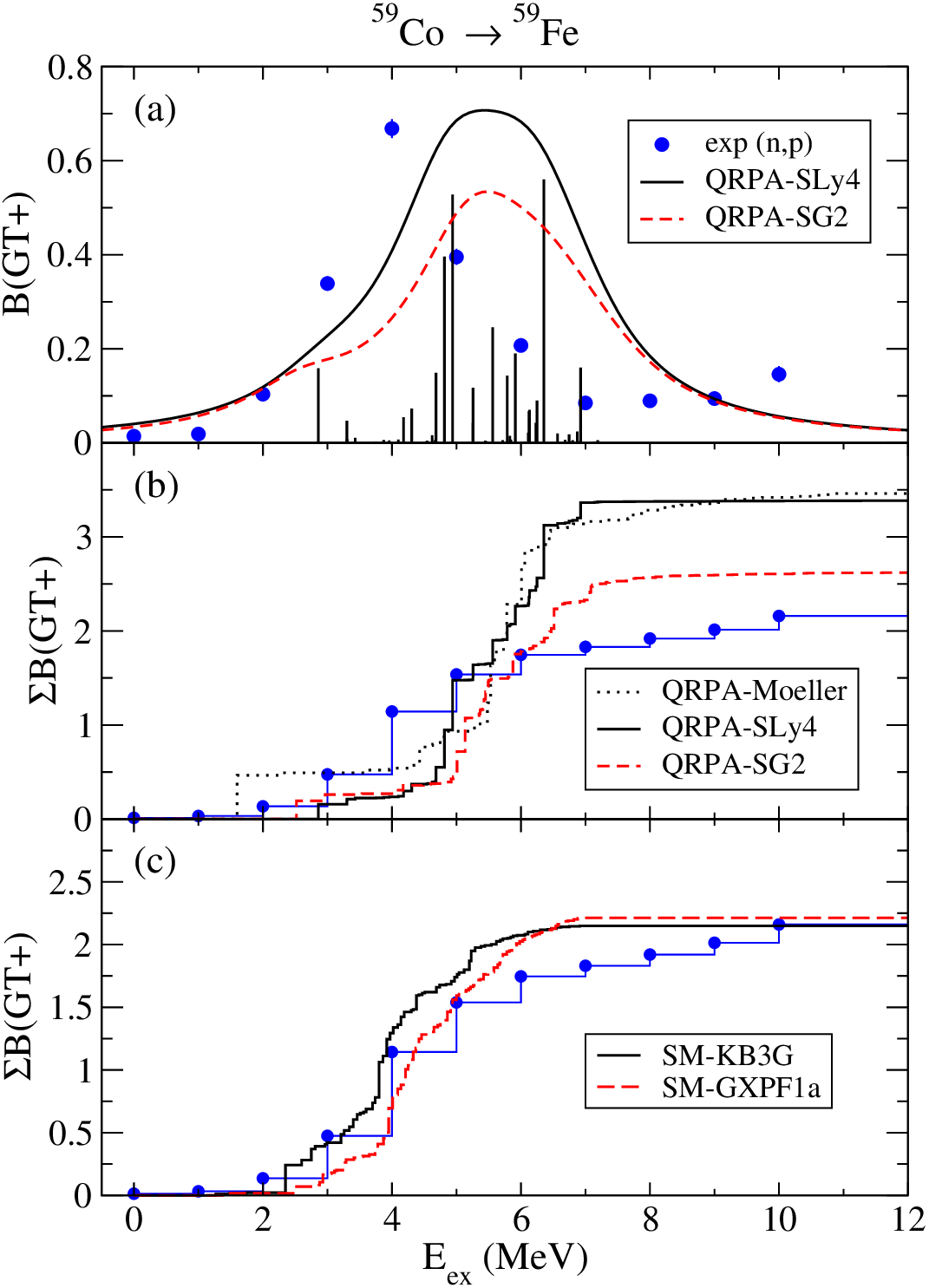}
\caption{(Color online) Same as in Fig. \ref{bgt_sc45}, but for the
$^{59}$Co to $^{59}$Fe transition.
Data are from $(n,p)$ reactions \cite{alford-93}.} 
\label{bgt_co59}
\end{figure}

\begin{figure}
\includegraphics[width=0.39\textwidth]{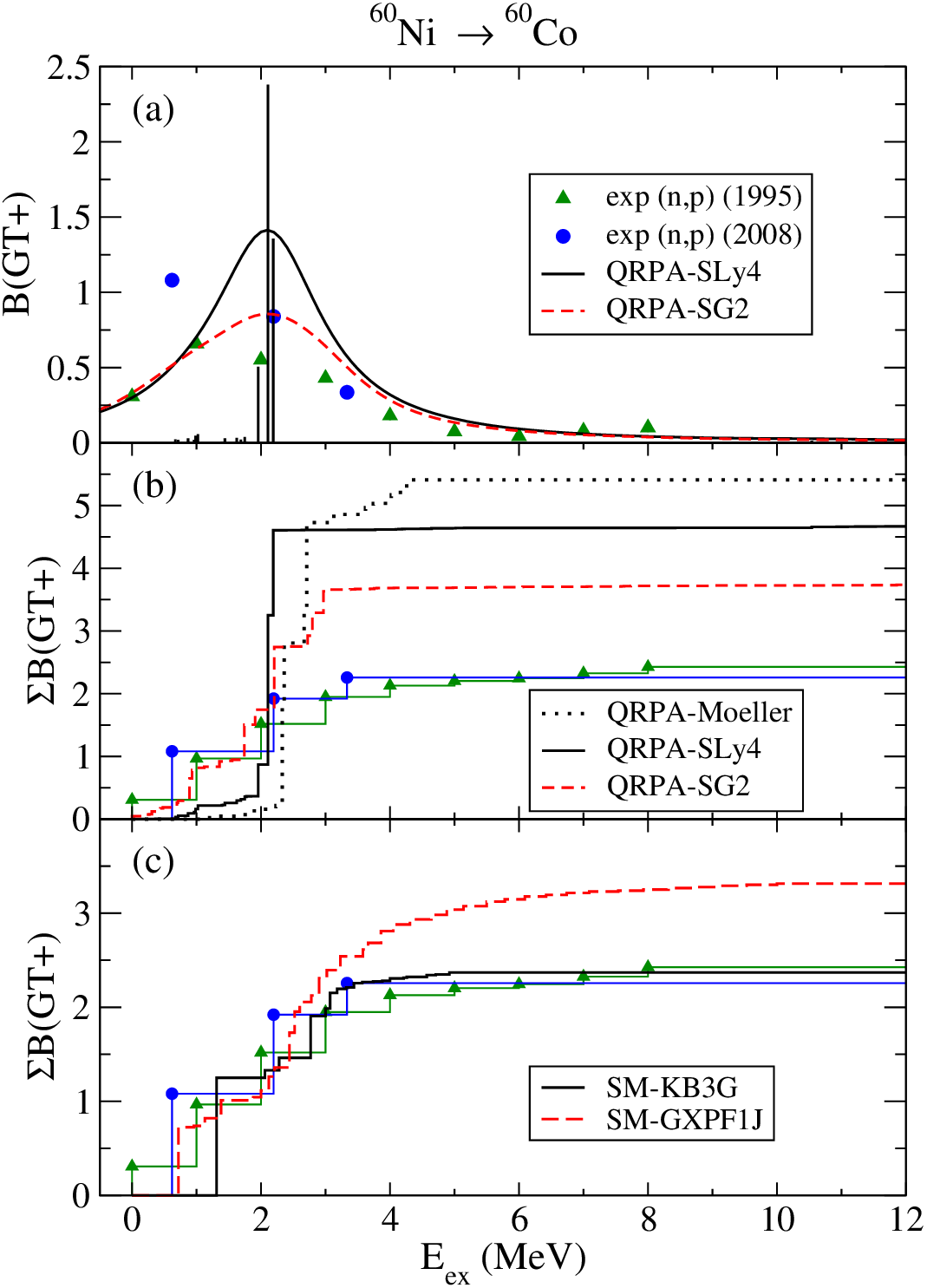}
\caption{(Color online) Same as in Fig. \ref{bgt_sc45}, but for the
$^{60}$Ni to $^{60}$Co transition.
Data are from $(n,p)$ reactions from Refs. \cite{williams-95,anantaraman-08}}.
\label{bgt_ni60}
\end{figure}

\begin{figure}
\includegraphics[width=0.39\textwidth]{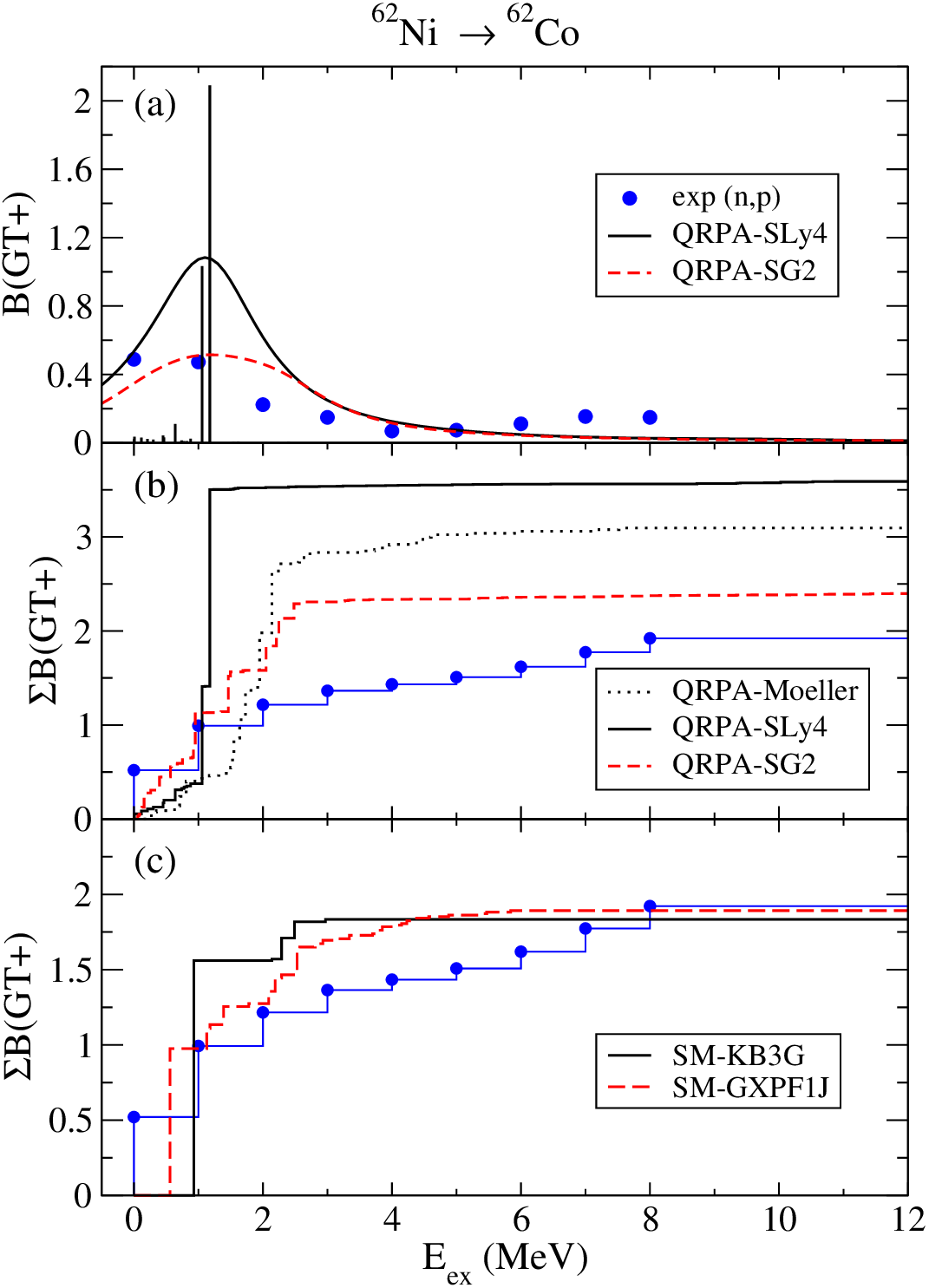}
\caption{(Color online) Same as in Fig. \ref{bgt_sc45}, but for the
$^{62}$Ni to $^{62}$Co transition.
Data are from $(n,p)$ reactions \cite{williams-95}.} 
\label{bgt_ni62}
\end{figure}

\begin{figure}
\includegraphics[width=0.39\textwidth]{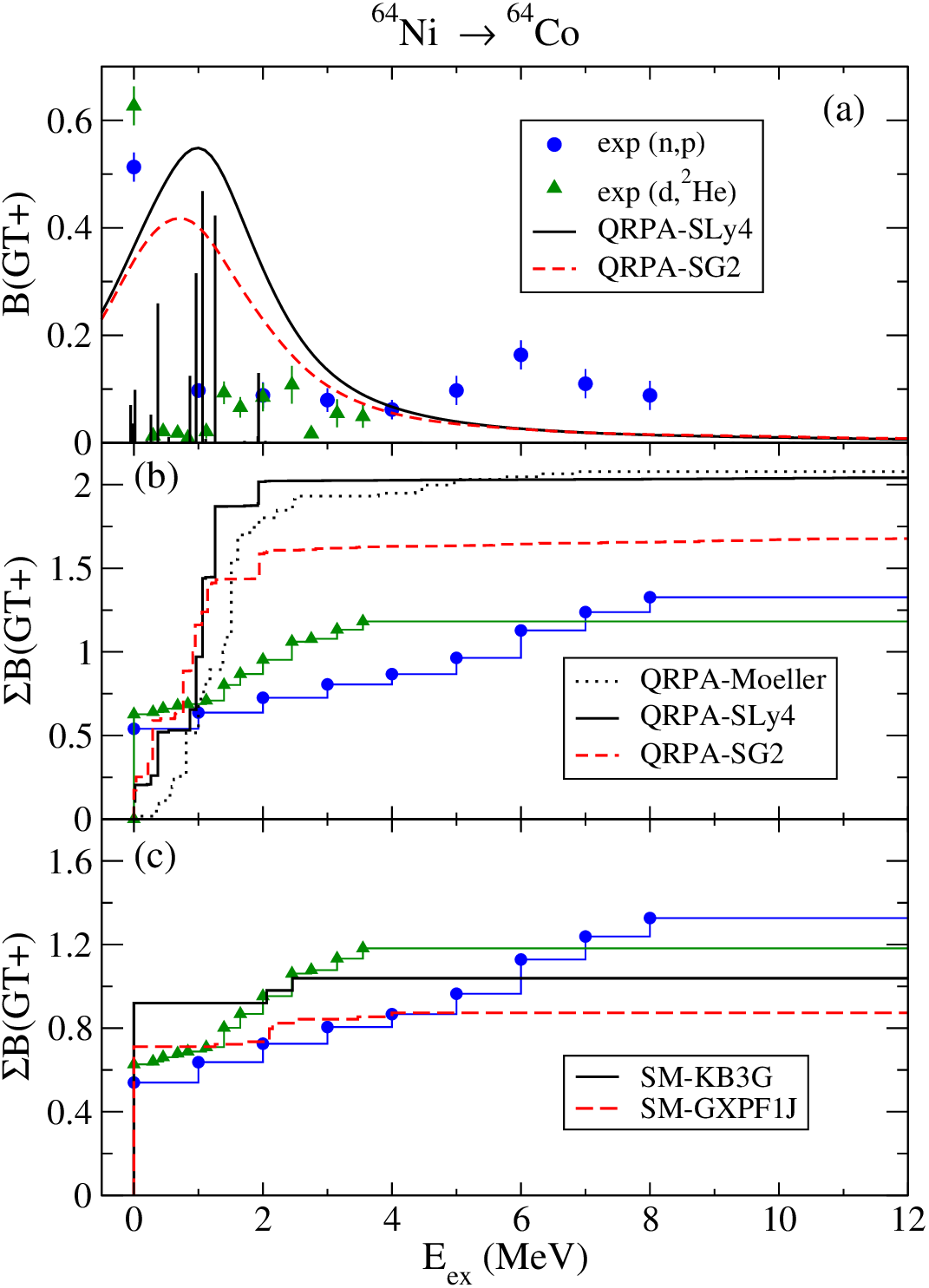}
\caption{(Color online) Same as in Fig. \ref{bgt_sc45}, but for the
$^{64}$Ni to $^{64}$Co transition.
Data are from $(n,p)$ \cite{williams-95} and $(d,^2{\rm He})$ 
\cite{popescu-07} reactions.} 
\label{bgt_ni64}
\end{figure}

\begin{figure}
\includegraphics[width=0.39\textwidth]{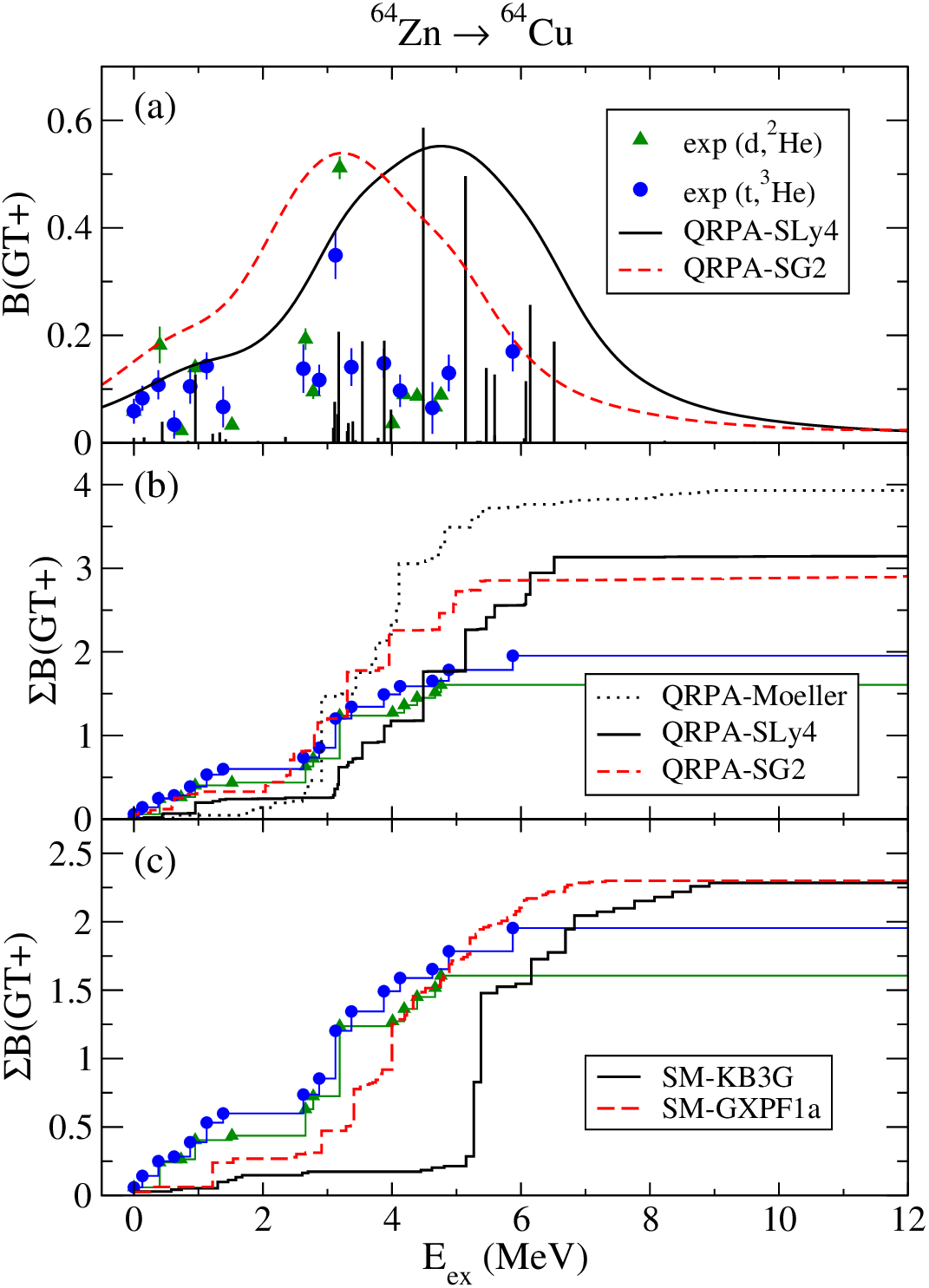}
\caption{(Color online) Same as in Fig. \ref{bgt_sc45}, but for the
$^{64}$Zn to $^{64}$Cu transition. Data are from 
$(d,^2{\rm He})$ \cite{grewe-08} and $(t,^3{\rm He})$ \cite{hitt-09}
reactions.} 
\label{bgt_zn64}
\end{figure}

\subsection{Gamow-Teller distributions}

In the next figures, we show the results obtained for the energy
distributions of the GT strength corresponding to the equilibrium
shapes for which we obtained minima in the potential energy curves. 
The GT strength is plotted versus the excitation
energy of the daughter nucleus $E_{ex}=E_f$ (MeV).

Figs. \ref{bgt_sc45}-\ref{bgt_zn64} contain the results for the
isotopes $^{45}$Sc, $^{48}$Ti, $^{51}$V, $^{54}$Fe, $^{55}$Mn, $^{56}$Fe,
$^{58}$Ni, $^{59}$Co, $^{60}$Ni, $^{62}$Ni, $^{64}$Ni, and $^{64}$Zn,
respectively.
This set of nuclei, which are the object of our study, have been 
chosen in accordance with the cases studied in Ref. \cite{cole-12}.

In the top panels (a) we show the experimental energy distributions and 
the individual GT strengths corresponding to the SLy4 interaction. 
We also show the continuous distributions from SLy4 and SG2 obtained by 
folding the strength with Breit-Wigner functions as it was done in 
Refs. \cite{sarri-03,radha-97,caurier-99}, so that the original discrete 
spectrum is transformed into a continuous profile. 
In the middle panels (b) we plot the measured energy distribution of the 
accumulated strength, which is compared to three QRPA calculations. They
are the QRPA-M\"oller calculations and the two calculations obtained from
our present formalism using the Skyrme forces SLy4 and SG2. Finally, in
the bottom panels (c) the comparison of the experimental distribution
is made with SM calculations with the interactions KB3G 
\cite{poves-01} and GXPF1a \cite{honma-02}, as given in Ref. \cite{cole-12}.

Fig. \ref{bgt_sc45} shows the GT strength distributions for the
$^{45}$Sc$(7/2^-)\rightarrow ^{45}$Ca extracted from $(n,p)$ charge-exchange
reactions \cite{alford-91}. Results are available in 1 MeV energy bins.
The measured strength distribution is compared to our QRPA results
in Fig. \ref{bgt_sc45}(a). The structure of the calculated distribution
agrees qualitatively with the data in the sense that the GT strength is 
concentrated around 7 MeV with small bumps at low energy. Both SG2 and 
SLy4 results are very similar. The total accumulated strength as a function of the 
excitation energy can be seen in Fig. \ref{bgt_sc45}(b), where the running 
sums are compared. Our calculations follow closely the experimental strength, 
while QRPA-M\"oller overestimates it. Similarly, the SM calculations shown 
in Fig. \ref{bgt_sc45}(c) compare reasonably well with the experiment with 
the strength practically concentrated in a single peak at 6 MeV and somewhat 
smaller total strength in the whole energy range.

Fig. \ref{bgt_ti48} contains the results for the 
$^{48}$Ti$(0^+)\rightarrow ^{48}$Sc transition. In this case the 
data are from $(n,p)$ \cite{yako-09} and $(d,^2{\rm He})$ \cite{rakers-04} 
reactions.
One should note that the strength measured in Ref. \cite{yako-09} would
contain an isovector spin monopole component, which is estimated to
contribute to about one third of the total strength. In the case
of the $(d,^2{\rm He})$ reaction, the high resolution achieved was 120 keV,
but the strength was extracted only up to an excitation energy of 5 MeV.
Both sets of data agree in the location of the peak of the strength
distribution at about 3 MeV, but disagree in other regions. Our QRPA
calculations show a two bump structure with peaks centered at about 4 and
6 MeV that resembles the profile of the $(n,p)$ data. The total strength
obtained lies between the strengths from  $(n,p)$ and $(d,^2{\rm He})$
reactions and is lower than QRPA-M\"oller. The SM calculations accumulate
the strength around 4 MeV containing less strength than QRPA.

Fig. \ref{bgt_v51} for the $^{51}$V$(7/2^-)\rightarrow ^{51}$Ti transition
contains data from $(n,p)$ \cite{alford-93} and $(d,^2{\rm He})$ \cite{baumer-03} 
reactions with 1 MeV and 110 keV resolution, respectively. Strengths from
the high resolution experiments were only extracted up to 6.5 MeV.
The main characteristic of these data is that the strength appears concentrated
at about 5 MeV excitation energies, a feature that is well reproduced
by our QRPA results although at somewhat higher energy.
The total strength measured in both experiments are similar and agrees 
quite well with our QRPA and with SM results. 

The results for the $^{54}$Fe$(0^+)\rightarrow ^{54}$Mn transition appear
in Fig. \ref{bgt_fe54}. In this case two sets of data from $(n,p)$ reactions
are available from Refs. \cite{ronnqvist-93,vetterli-87}.
Our QRPA results produce a bump centered at about 5 MeV which is a little
bit displaced to higher energies with respect to the experiment. The total
strengths in QRPA are also above the measured strength, which is better
reproduced by the SM calculations.

In the case of the $^{55}$Mn$(5/2^-) \rightarrow ^{55}$Cr transition in
Fig. \ref{bgt_mn55} there are data extracted from $(n,p)$ reactions 
\cite{elkateb-94}. The broad peak observed experimentally centered at
about 4 MeV is well reproduced in our QRPA calculations. Also the total
GT strength is in this case well accounted for by all the models.

Fig. \ref{bgt_fe56} shows the results corresponding to the 
$^{56}$Fe$(0^+)\rightarrow ^{56}$Mn transition.
Data are from $(n,p)$ reactions \cite{ronnqvist-93,elkateb-94}.
Both sets of data show a concentration of the GT strength between
1 and 4 MeV in agreement with the calculations. However, the
strength observed beyond 6 MeV is not found in any of the
calculations presented. The total strength measured is slightly
overestimated (underestimated) by QRPA (SM) calculations.

The distribution of the GT strength corresponding to the transition
$^{58}$Ni$(0^+)\rightarrow ^{58}$Co is shown in Fig. \ref{bgt_ni58}.
Data are from $(n,p)$ \cite{elkateb-94}, $(d,^2{\rm He})$ 
\cite{hagemann-04}, and $(t,^3{\rm He})$ \cite{cole-06} reactions
with energy resolutions of 1.2 MeV, 130 keV, and 250 keV, respectively.
Our QRPA calculations produce a strength distribution sharply 
concentrated between 3 and 4 MeV that contains practically all the
GT strength observed. The SM results are more fragmented in better
agreement with experiment. Contrary to previous cases, we use here
and in the rest of Ni isotopes, the SM results obtained with the 
improved interaction GXPF1J \cite{suzuki-09}.

Fig. \ref{bgt_co59} shows the results for the 
$^{59}$Co$(7/2^-)\rightarrow ^{59}$Fe transition.
Data in this case have been obtained from $(n,p)$ 
reactions \cite{alford-93} with about 1 MeV energy resolution.
QRPA results show one peak structure centered around 5 MeV, which
seems to be displaced 1 MeV to higher energy with respect to
the experimental distribution. The total GT strength is somewhat
larger than experiment. On the other hand SM results agree
quite well with the experiment.

In the case of the $^{60}$Ni$(0^+)\rightarrow  ^{60}$Co transition shown
in Fig. \ref{bgt_ni60}, the data are from $(n,p)$ reactions 
\cite{williams-95} and from the reanalysis performed in Ref.
\cite{anantaraman-08}. As in the case of $^{58}$Ni, whereas the
experimental strength appears fragmented below 3 MeV, QRPA calculations
produce a sharp transition with practically all the strength at 2 MeV.
All the calculations except SM-KB3G overestimate the total strength
measured.

The case of the $^{62}$Ni$(0^+)\rightarrow  ^{62}$Co transition
(Fig. \ref{bgt_ni62}) is very similar to the previous case for
$^{60}$Ni. The data from $(n,p)$ reactions \cite{williams-95}
show that practically all the strength is contained below 2 MeV,
which is compatible with the structure of the GT strength distributions
obtained from the theoretical models. However, while QRPA calculations
overestimate the total strength, SM results agree very nicely with
experiment.

For the $^{64}$Ni$(0^+)\rightarrow  ^{64}$Co transition in 
Fig. \ref{bgt_ni64} we have data not only from $(n,p)$ reactions
\cite{williams-95} as in the previous Ni isotopes, but also 
data up to 4 MeV from
$(d,^2{\rm He})$ \cite{popescu-07} reactions at a much higher 
energy resolution (110 keV). Most of the GT strength is observed
in the ground state to ground state transition, while in QRPA 
the strength is fragmented below 2 MeV. The calculated strength
is overestimated (underestimated) by QRPA (SM) calculations. 

In the last example, we see in Fig. \ref{bgt_zn64} the results for the
$^{64}$Zn$(0^+)\rightarrow ^{64}$Cu transition. Data are from 
$(d,^2{\rm He})$ \cite{grewe-08} and $(t,^3{\rm He})$ \cite{hitt-09}
reactions achieving 115 keV and 280 keV energy resolution, respectively.
The QRPA results show a bump centered at an excitation energy of 3 MeV 
(5 MeV) when using SG2 (SLy4) interaction. SG2 results reproduce better
the experimental results, which show a strong peak at 3 MeV.
The calculated total strength, including results from SM calculations, 
overestimates always the experiment. Results from SM-KB3G produce a strong
peak at 5 MeV at variance with experiment.

Summarizing this section we can say that the observed fragmentation
of the experimental GT$^+$ strength distributions over many states,
as well as the centroids and widths of the distributions are 
reasonably well described by the calculations. In the experimental
distributions, especially in those extracted from ($n,p$) reactions, 
there is a tendency to build up a second peak beyond 
$\sim6$ MeV, which is not reproduced in the calculations. In 
general, QRPA produce more strength at higher energy than SM 
because of the higher N-shell mixing contained in QRPA.
All in all, the present QRPA calculations based on the deformed Skyrme 
HF+BCS+QRPA described in Sec. \ref{form} are in general of comparable 
quality to SM calculations.
These results serve to refine the findings in Ref. \cite{cole-12},
where it was concluded that 
QRPA calculations based on Ref. \cite{krumlinde-84} produce
systematically much larger deviations from the data
than SM calculations. We have shown in this work that this is not 
necessarily the case for all QRPA type calculations.

\clearpage

\begin{figure}
\includegraphics[width=0.40\textwidth]{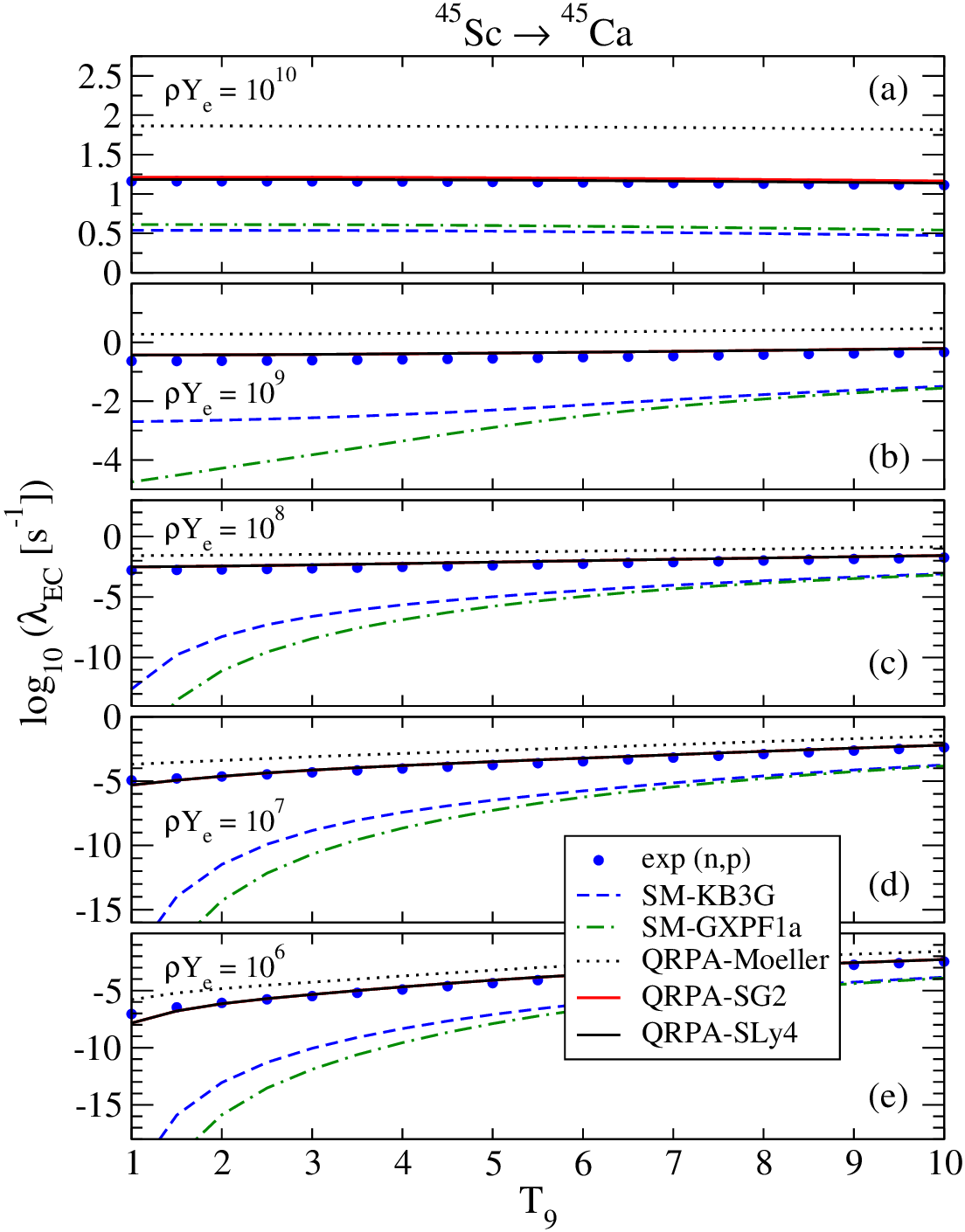}
\caption{(Color online) Electron-capture rates for $^{45}$Sc obtained 
from the experimental GT strength distributions and from different shell 
model and QRPA calculations as a function of the temperature $T_9$ (GK) 
for densities $\rho Y_e=10^{10},\ 10^9,\ 10^8,\ 10^7,\ 10^6$ mol/cm$^3$ 
in panels (a), (b), (c), (d), and (e), respectively.}
\label{rate_sc45}
\end{figure}

\begin{figure}
\includegraphics[width=0.40\textwidth]{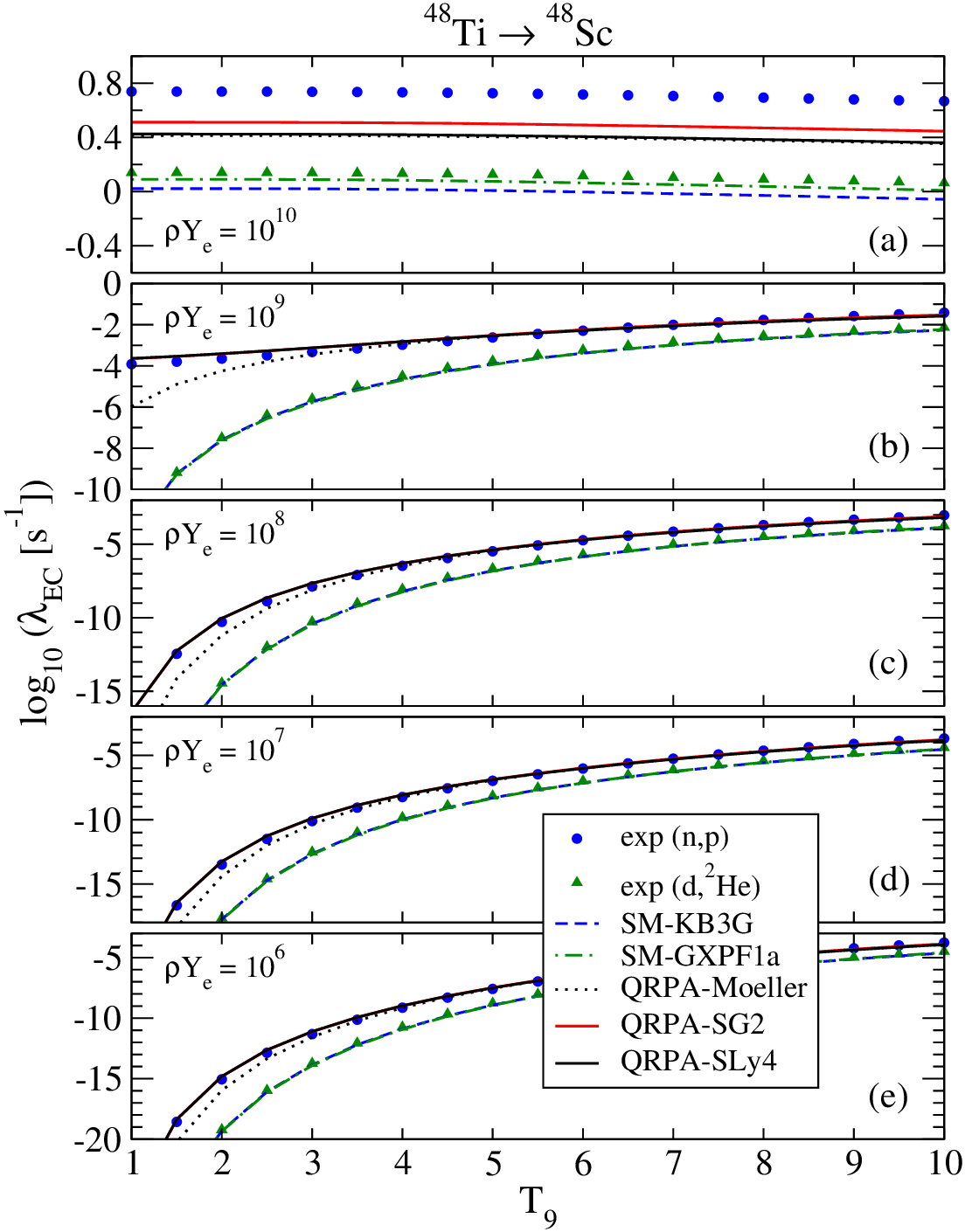}
\caption{(Color online) Same as in Fig. \ref{rate_sc45}, but for $^{48}$Ti.}
\label{rate_ti48}
\end{figure}

\begin{figure}
\includegraphics[width=0.40\textwidth]{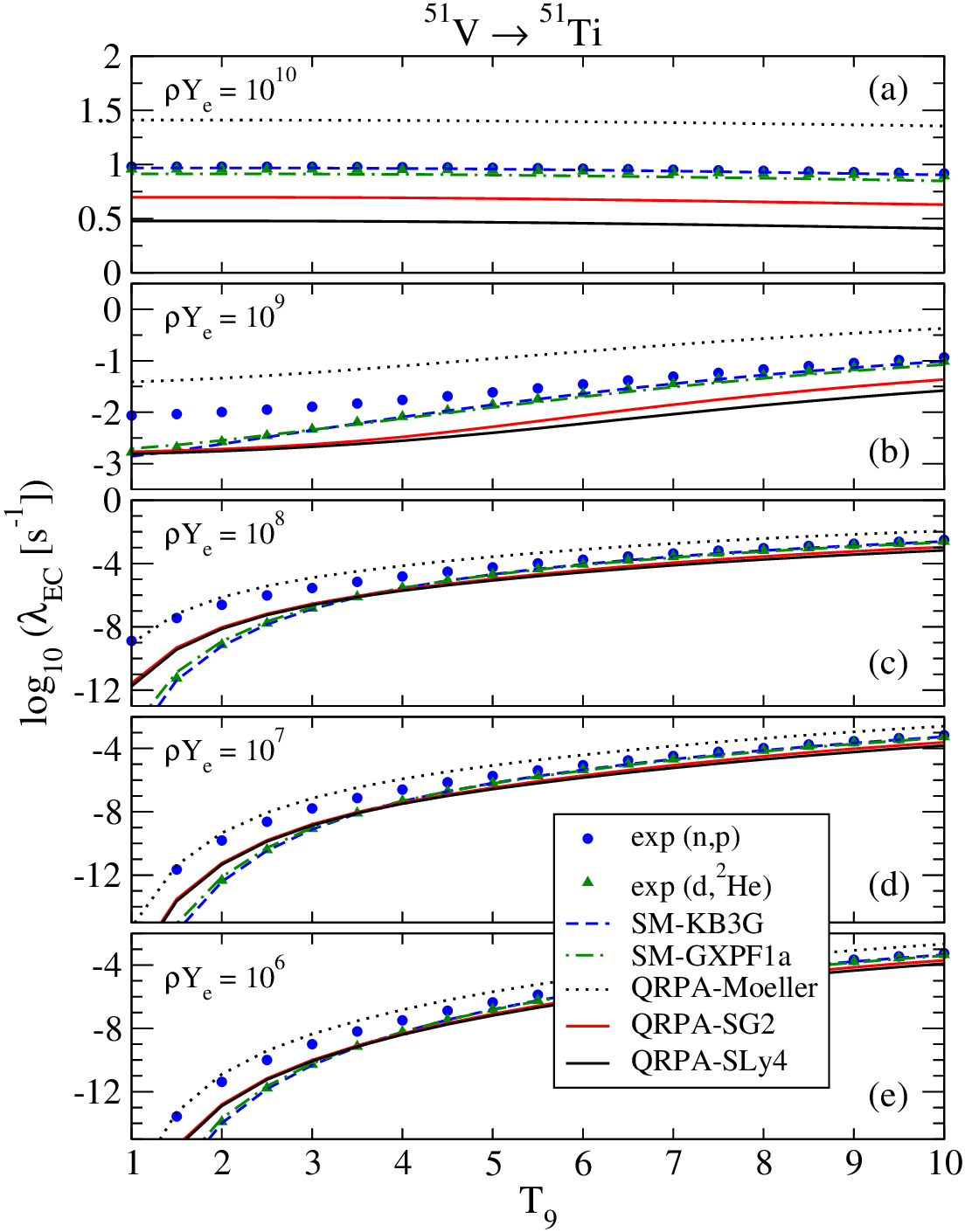}
\caption{(Color online) Same as in Fig. \ref{rate_sc45}, but for $^{51}$V.}
\label{rate_v51}
\end{figure}

\begin{figure}
\includegraphics[width=0.40\textwidth]{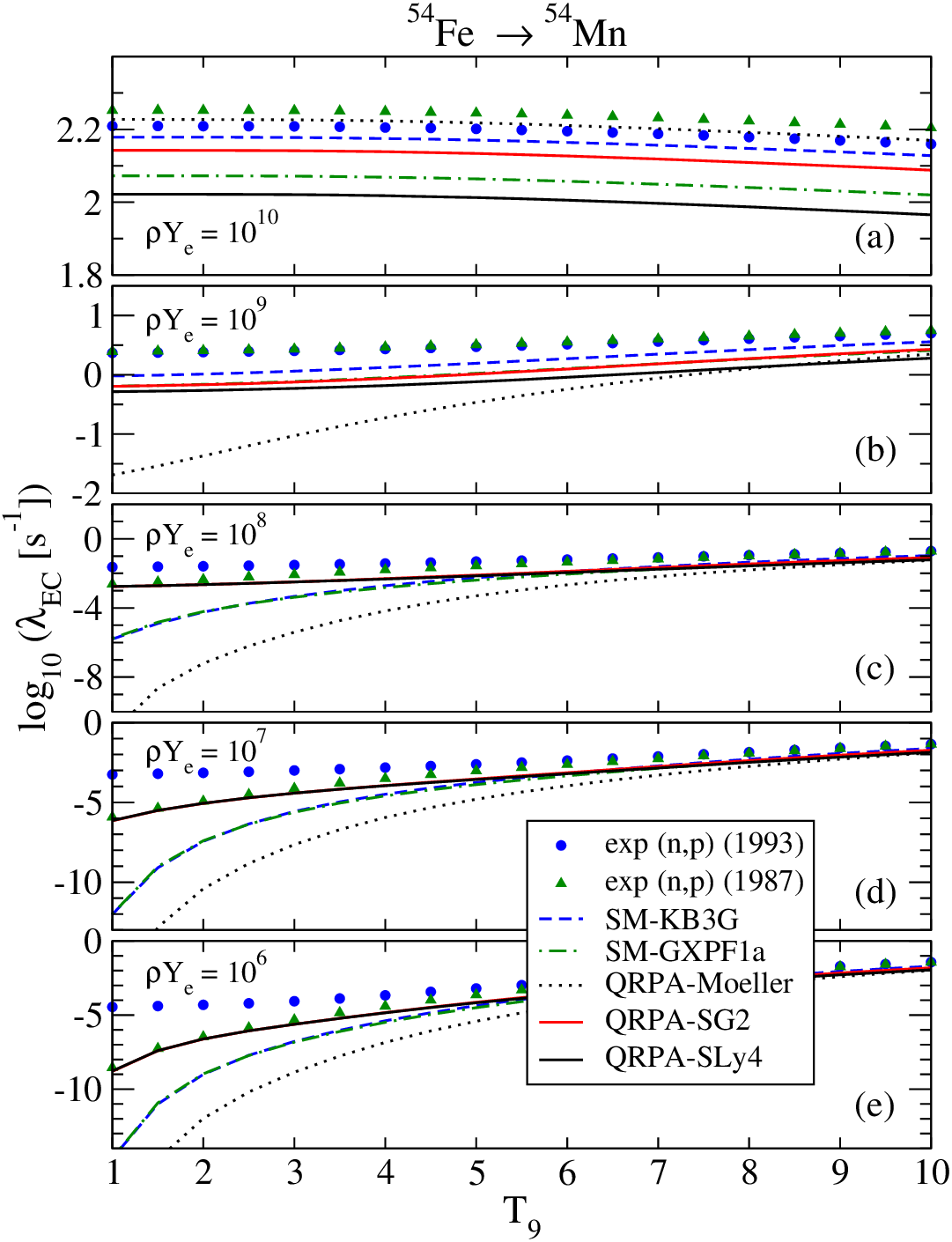}
\caption{(Color online) Same as in Fig. \ref{rate_sc45}, but for $^{54}$Fe.}
\label{rate_fe54}
\end{figure}

\begin{figure}
\includegraphics[width=0.40\textwidth]{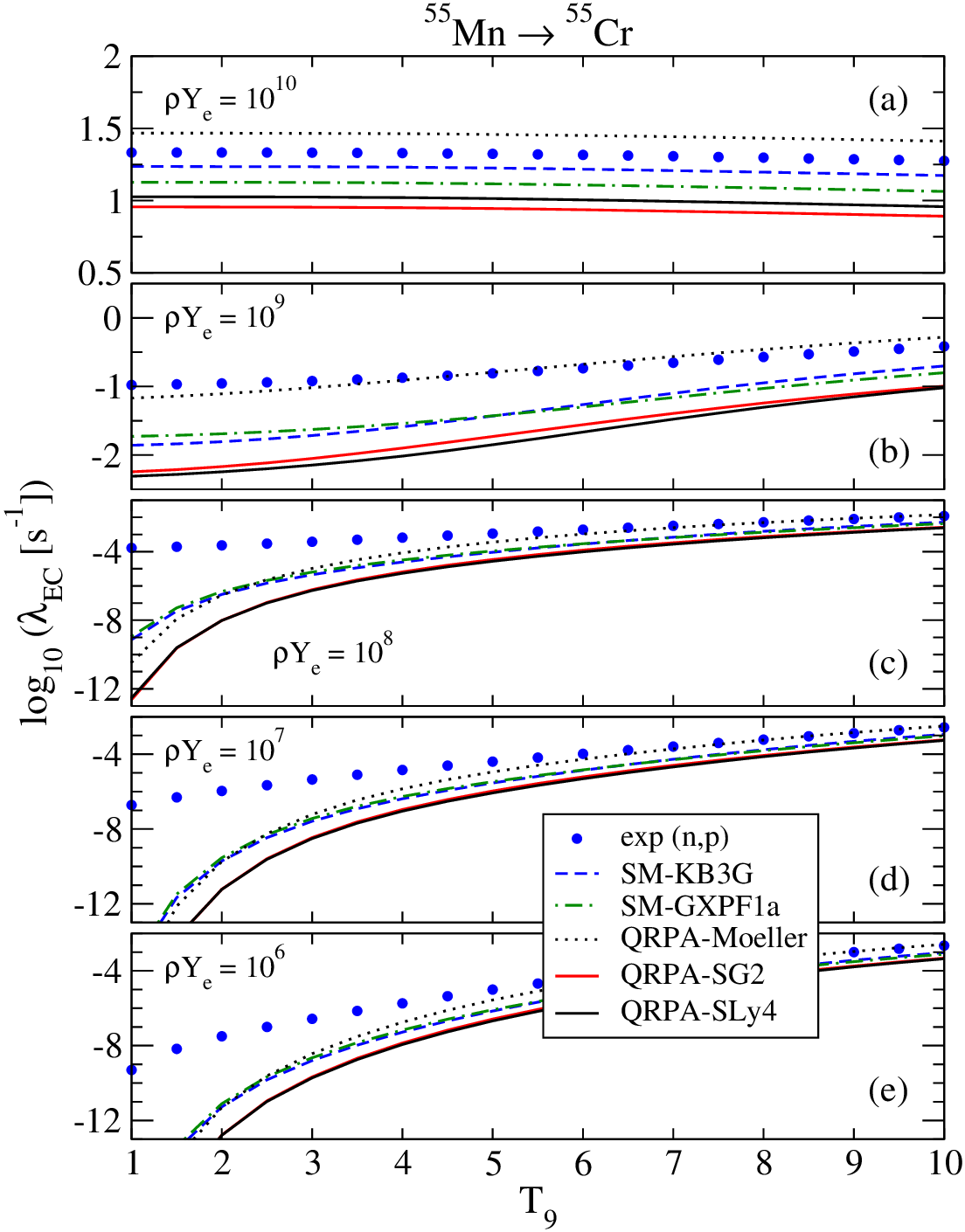}
\caption{(Color online) Same as in Fig. \ref{rate_sc45}, but for $^{55}$Mn.}
\label{rate_mn55}
\end{figure}

\begin{figure}
\includegraphics[width=0.40\textwidth]{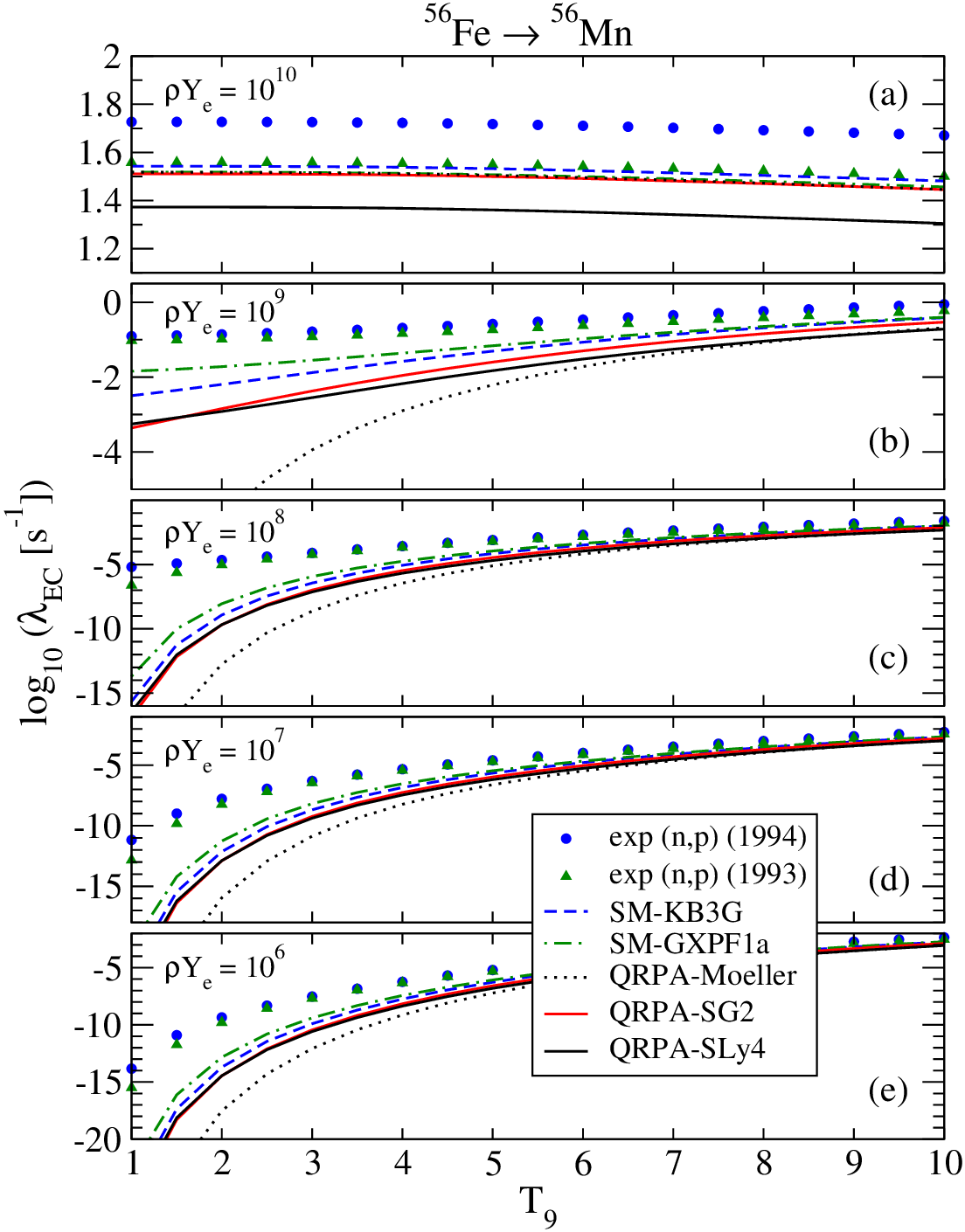}
\caption{(Color online) Same as in Fig. \ref{rate_sc45}, but for $^{56}$Fe.}
\label{rate_fe56}
\end{figure}

\begin{figure}
\includegraphics[width=0.40\textwidth]{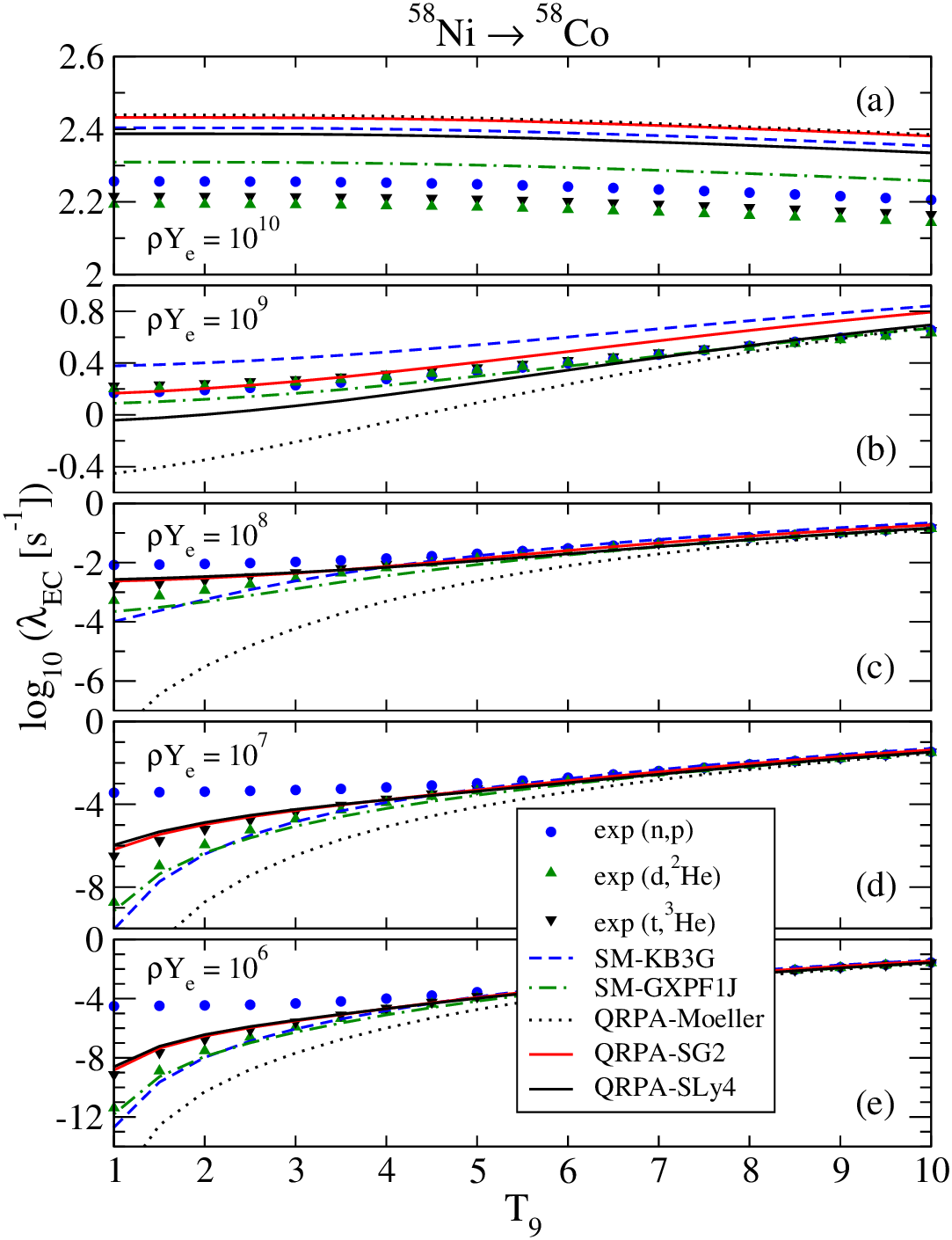}
\caption{(Color online) Same as in Fig. \ref{rate_sc45}, but for $^{58}$Ni.}
\label{rate_ni58}
\end{figure}

\begin{figure}
\includegraphics[width=0.40\textwidth]{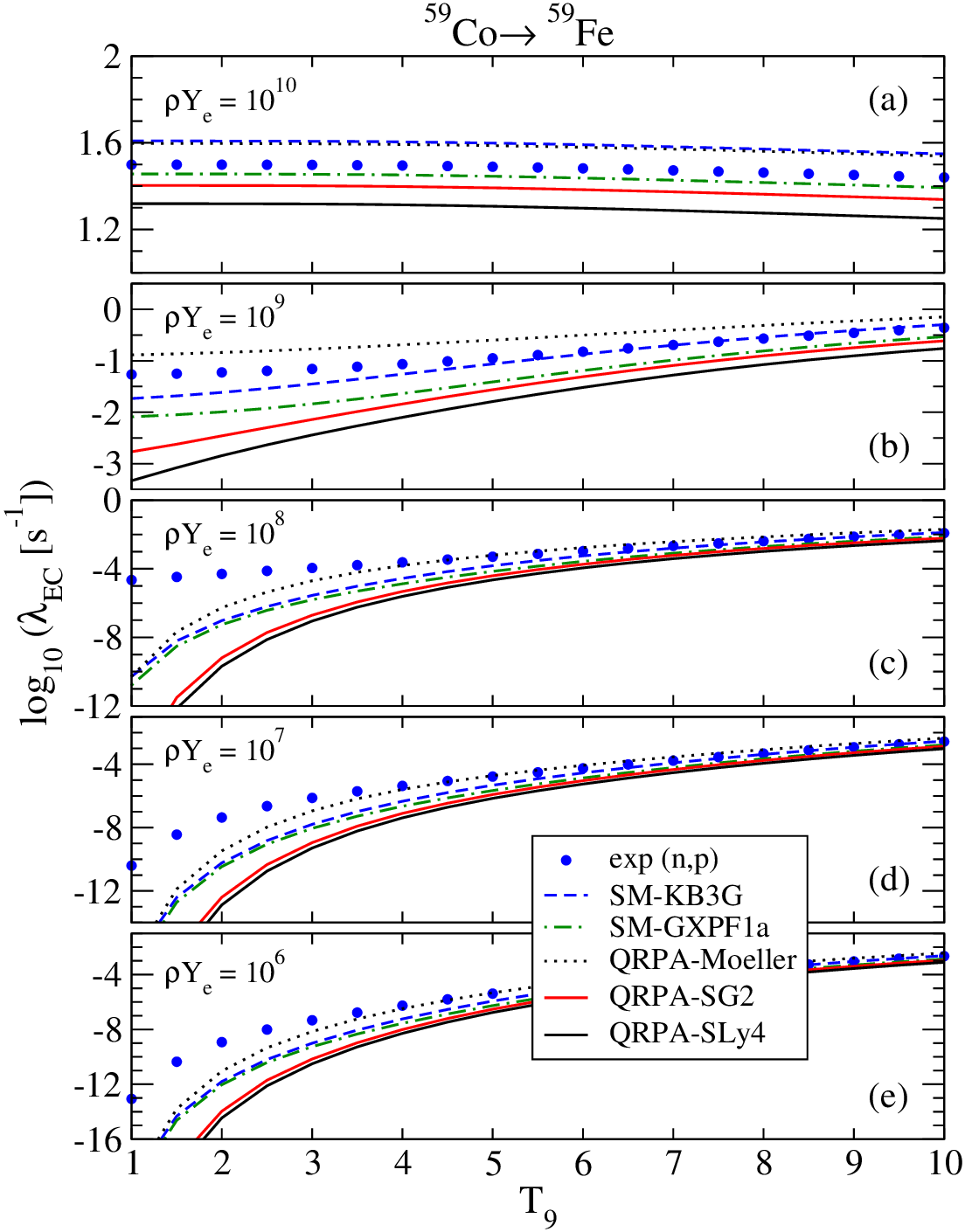}
\caption{(Color online) Same as in Fig. \ref{rate_sc45}, but for $^{59}$Co.}
\label{rate_co59}
\end{figure}

\begin{figure}
\includegraphics[width=0.40\textwidth]{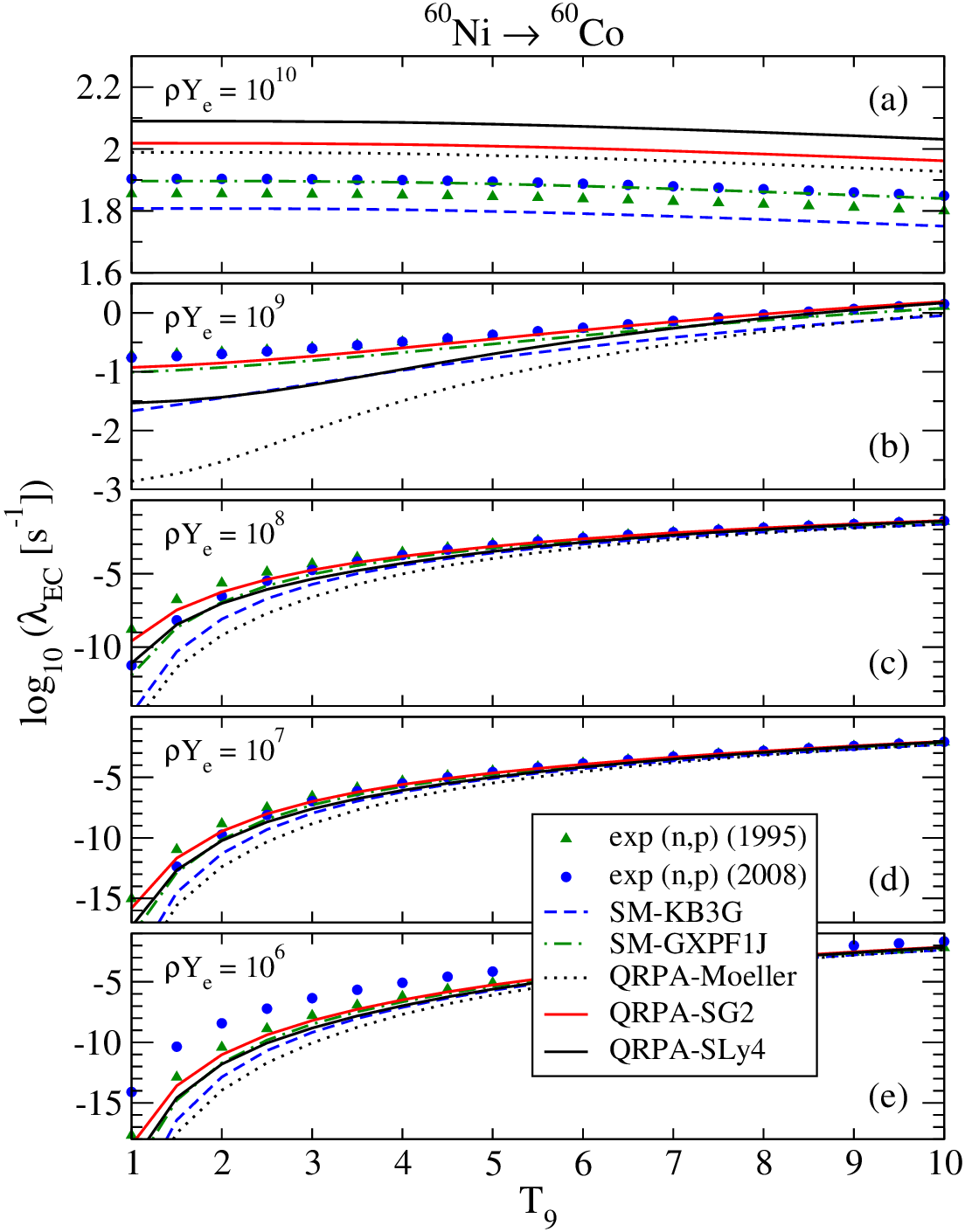}
\caption{(Color online) Same as in Fig. \ref{rate_sc45}, but for $^{60}$Ni.}
\label{rate_ni60}
\end{figure}

\begin{figure}
\includegraphics[width=0.40\textwidth]{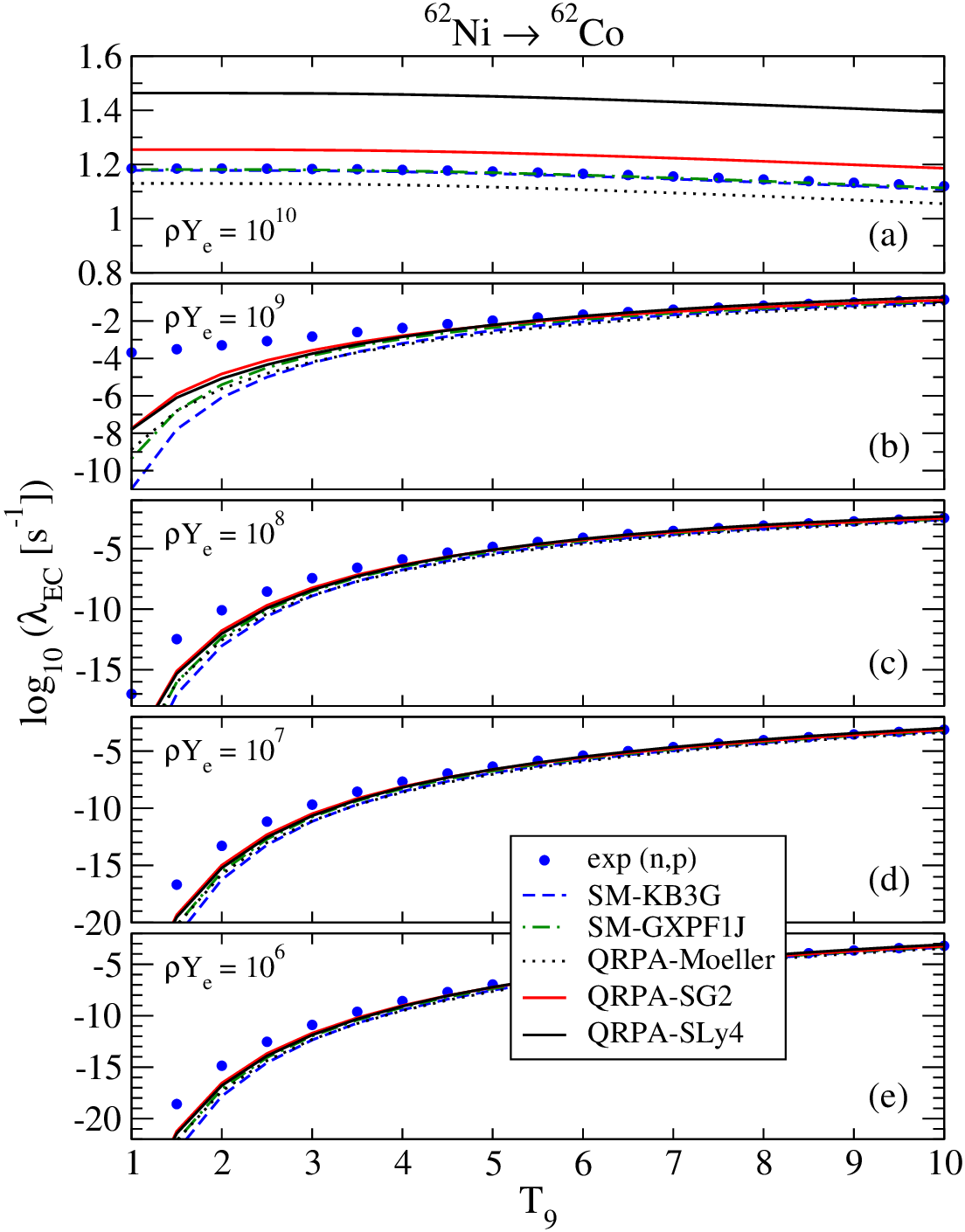}
\caption{(Color online) Same as in Fig. \ref{rate_sc45}, but for $^{62}$Ni.}
\label{rate_ni62}
\end{figure}

\begin{figure}
\includegraphics[width=0.40\textwidth]{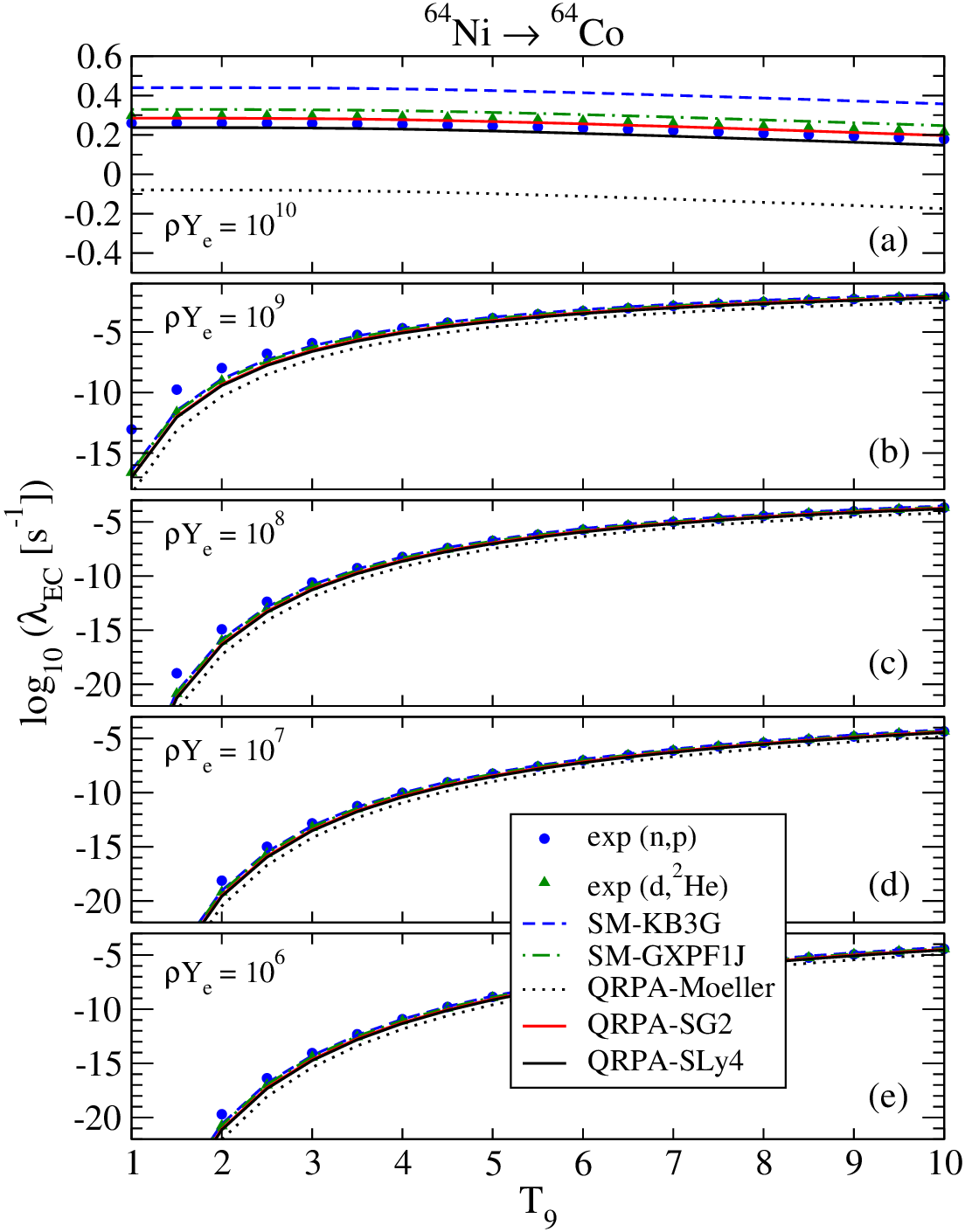}
\caption{(Color online) Same as in Fig. \ref{rate_sc45}, but for $^{64}$Ni.}
\label{rate_ni64}
\end{figure}

\begin{figure}
\includegraphics[width=0.40\textwidth]{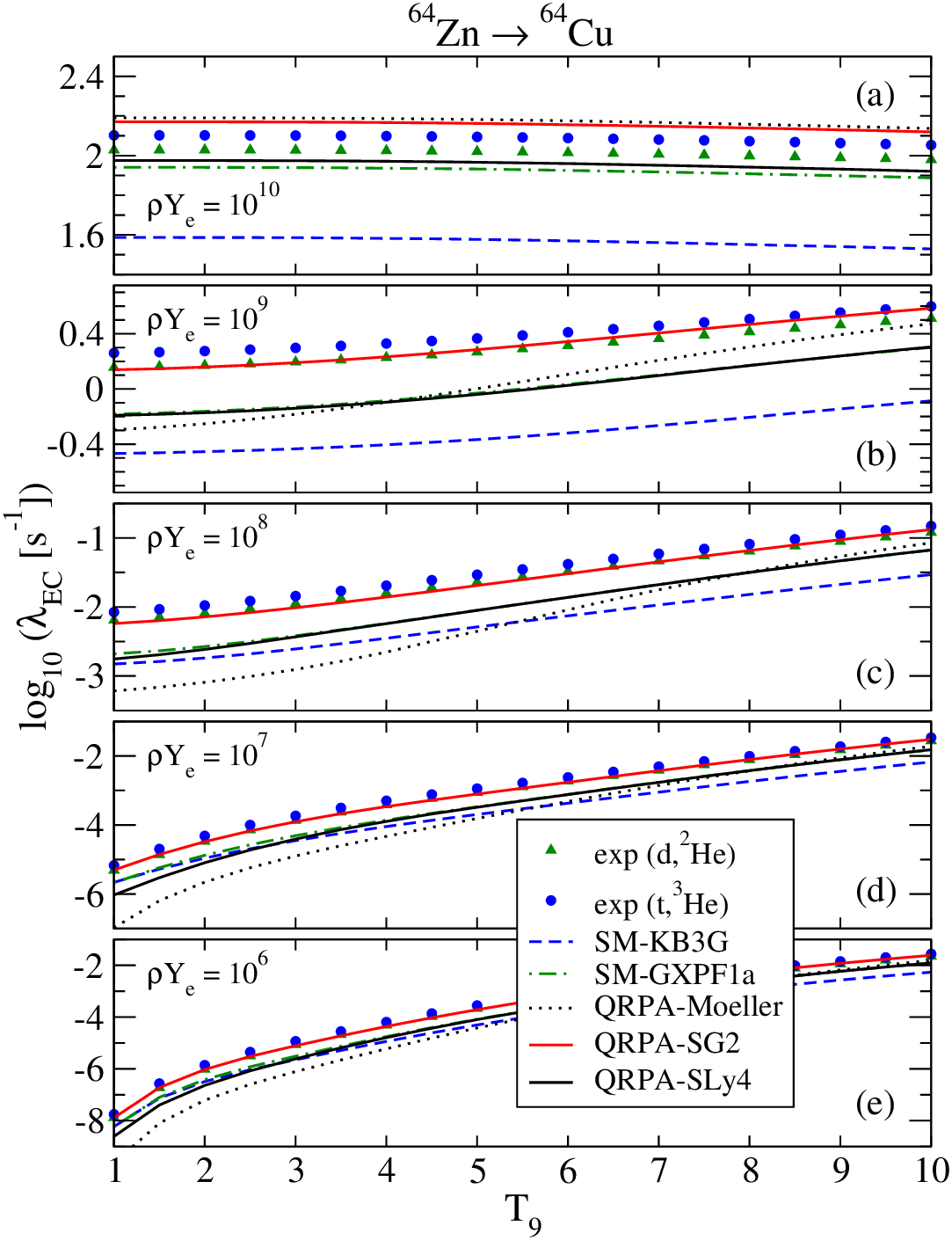}
\caption{(Color online) Same as in Fig. \ref{rate_sc45}, but for $^{64}$Zn.}
\label{rate_zn64}
\end{figure}

\subsection{Stellar weak decay rates}

In the following figures (Figs. \ref{rate_sc45}-\ref{rate_zn64}) we present 
the EC rates of the selected {\it pf}-shell nuclei as a function of the 
temperature and for various densities. 
The range of $T$ considered varies from $T_9=1$ up to $T_9=10$, 
whereas the range in $\rho Y_e$ varies from $\rho Y_e=10^6$
mol/cm$^3$ up to  $\rho Y_e=10^{10}$ mol/cm$^3$. This grid of $\rho$ and $T$ 
includes those ranges relevant for astrophysical scenarios related to the 
silicon-burning stage in a presupernovae star \cite{aufderheide-94}
($\rho Y_e=10^{7}$ mol/cm$^3$ and  $T_9=3$), as well as scenarios related to 
pre-collapse of the core \cite{hix-03} and thermonuclear runaway type Ia 
supernovae \cite{iwamoto-99} ($\rho Y_e=10^{9}$ mol/cm$^3$ and  $T_9=10$).
In each figure from Fig. \ref{rate_sc45} to Fig. \ref{rate_zn64} 
we show the EC rates obtained from the experimental GT strength distributions 
and from different SM (KB3G and GXPF1) and QRPA (SLy4, SG2, and M\"oller) 
calculations.

Concerning the experimental EC rates $\lambda_{\rm EC, exp}$ in Figs.
\ref{rate_sc45}-\ref{rate_zn64}, one should keep in mind that these 
quantities are not necessarily the actual rates in stellar scenarios 
at high $\rho$ and $T$. $\lambda_{\rm EC, exp}$ is indeed the rate calculated 
from the experimental GT strength distribution extracted from charge-exchange 
reactions. Besides the intrinsic uncertainties in the extraction of the GT 
strength due to several causes like the global normalization of the unit 
cross section or to possible interferences caused by the tensor component 
of the interaction, which are typically estimated to be a 10-20\% effect, 
one has to consider other sources of ignorance that makes 
$\lambda_{\rm EC, exp}$ to be different from the EC rates in stellar scenarios. 
First, the GT strength is only measured up to some 
excitation energy and therefore, $\lambda_{\rm EC, exp}$ does not include 
possible contributions from transitions beyond the measured energy range 
that could have an effect, especially at high $T$ and $\rho$.
Secondly, even when the GT strength distribution is perfectly determined 
from charge-exchange reactions in the laboratory under terrestrial conditions, 
this is not sufficient to determine the EC rates in stellar scenarios where 
ECs can occur on excited states of the parent nucleus that become thermally 
populated at high $T$.
Although these effects are expected to be very small for the densities and
temperatures considered, one should keep in mind the real significance
of $\lambda_{\rm EC, exp}$.

Some general comments about the sensitivity of the EC rates to $(\rho ,T)$, 
to the Fermi and $Q_{EC}$ energies, and to the GT distribution are in order 
to understand better the EC rates. 
Values for $Q_{EC}$ and Fermi energies are given in Table I and Table II, 
respectively. Since the Fermi energy increases with
the density, it is expected that the EC rates at low densities are mainly 
sensitive to the GT strength of states at low excitation energies in the 
daughter nucleus.
This is especially true at low $T$, where the shape of the electron energy
distribution, $S_e$, has a sharp surface at the Fermi energy. When $T$ 
raises, the shape of $S_e$ is smeared out and thus, GT transitions at higher 
excitation energies might contribute to the rates even at low densities.
When $\rho$ increases the rates also increase because the Fermi energy of the 
electrons is larger and larger allowing to reach higher excitation energies 
in the daughter nucleus and thus making the GT strength at these energies 
contribute to the rates in a more significant way.
The magnitude of this general effect is different in each nucleus because of 
the different $Q_{EC}$ values that force the electron energy to be large enough
to overcome it.
When increasing $T$, the rates in general increase because the diffuseness of 
the electron energy distributions makes it possible that more excited states 
at higher energy can be reached.

On the other hand the role of the $Q_{if}$ energy in Eq. (\ref{qif}) is also
very important. It determines the lower integration limit in the EC phase 
factor (\ref{phiec}). Then, the energy of the available electrons must
overcome this value to be captured or in other words, the Fermi energy
has to be larger than $Q_{if}$. The argument is strictly valid at $T=0$ 
where $S_e$ changes abruptly from one to zero at the Fermi energy. When $T$ 
increases, $S_e$ is smeared out and ECs are possible even for Fermi energies 
lower than $Q_{if}$. In general, the larger the $Q_{EC}$ energy (more negative) 
the lower the EC rates. This effect will be accentuated at low densities where
the Fermi energy is small.
Similarly, nuclei with large (negative) $Q_{EC}$ values (i.e., $^{48}$Ti, 
$^{56}$Fe, $^{62}$Ni, $^{64}$Ni) are mostly sensitive to the GT strength of the
ground and lowest excited states and the opposite is true for the nuclei 
with small $Q_{EC}$ values ($^{45}$Sc, $^{54}$Fe, $^{58}$Ni, $^{64}$Zn).

It is also worth comparing the rates in the Ni isotopes that exhibit
different values of their $Q_{EC}$ energies (see Table I).
The EC rates in $^{64}$Ni ($Q_{EC}=-7.307$ MeV) are almost independent 
of the density for values between $10^6$ and $10^8$. They start growing
at $10^9$ and become very large in comparison at $10^{10}$. 
This effect is related to the fact that  the Fermi energy is about
11 MeV (see Table II) at the highest density, which is enough to surpass 
the large $Q_{EC}$ value.
We find a similar situation in the case of $^{62}$Ni ($Q_{EC}=-5.315$ MeV)
and to a less extent in $^{60}$Ni ($Q_{EC}=-2.823$ MeV). Finally,
in $^{58}$Ni ($Q_{EC}=-0.382$ MeV) the rates increase more steadily
because the Fermi energy exceeds the small $Q_{EC}$ much sooner.
Nuclei with large negative $Q_{EC}$ energies such as $^{48}$Ti and $^{56}$Fe show 
similarities with the rates in $^{64}$Ni, whereas nuclei with small negative
$Q_{EC}$ energies like $^{45}$Sc, $^{54}$Fe, and $^{64}$Zn exhibit 
similarities with the rates in $^{58}$Ni.

We should also mention that in all the examples studied, the rates are 
practically independent of $T$ at $\rho Y_e=10^{10}$ mol/cm$^3$ and only 
a model dependence is apparent. This is a direct consequence of the large 
Fermi energy (about 11 MeV) at any $T$ that makes the rates sensitive to
the GT strength at all the excitation energies.


\begin{table*}[ht]
\begin{center}
\caption{ $Q_{EC}$ values [MeV] from Eq. (\ref{qec}) using experimental 
nuclear masses from Ref. \cite{ensdf}.  }
\vskip 0.5cm
\begin{tabular}{cccccccccccc}
\hline \hline \\
  $^{45}$Sc &  $^{48}$Ti & $^{51}$V  & $^{54}$Fe & $^{55}$Mn & $^{56}$Fe &
  $^{58}$Ni &  $^{59}$Co & $^{60}$Ni & $^{62}$Ni & $^{64}$Ni & $^{64}$Zn \\ \\
 -0.258 & -3.990 & -2.472 & -0.697 & -2.603 & -3.696 & -0.382 &
 -1.565 & -2.823 & -5.315 & -7.307 & -0.580 \\ 
\hline \hline
\end{tabular}
\end{center}
\label{tableqec}
\end{table*}

Now we discuss the model dependence of the EC rates. As we have already 
mentioned, as a general rule the EC rates at low $\rho$ and $T$ will be 
more sensitive to the GT strength at low excitation energies, especially 
for nuclei with large negative $Q_{EC}$ energies.
On the other hand, the rates at high $\rho$ and $T$ will depend 
more on the global structure of the GT distribution. This type of
correlations can be seen to some extent in all the figures from
Figs \ref{rate_sc45} to Fig. \ref{rate_zn64}.

In the case of $^{45}$Sc (Fig. \ref{rate_sc45}) the small value of 
$Q_{EC}$ makes the EC rates very sensitive to the nuclear model because
the GT distribution is to a large extent involved up to large excitation
energies. We can see that we always get the rates from QRPA-M\"oller 
larger than the rates from QRPA-SLy4 and QRPA-SG2, and larger than
the SM ones. This is clearly correlated to the GT strength distributions
in Fig. \ref{bgt_sc45}, where we observe the same ordering in the GT strength.
In particular, the rates from our QRPA calculations are very close to
the experimental ones, while QRPA-M\"oller gives larger rates and SM produce 
rates clearly above. This is especially noticeable at low $\rho$ and $T$ 
due to the absence of strength between 0-6 MeV in the SM calculations.

In the case of  $^{48}$Ti (Fig. \ref{rate_ti48}) with a relatively large 
$Q_{EC}$ value, we can see that the experimental rates
from $(n,p)$ are larger than the corresponding rates for $(d,^2{\rm He})$
in accordance with the GT strengths in Fig. \ref{bgt_ti48}. While
the QRPA calculations agree better with the rates from $(n,p)$, the SM
calculations agree better with the rates from  $(d,^2{\rm He})$.
In the case of $^{51}$V (Fig. \ref{rate_v51}) all the models are quite 
similar with the exception of QRPA-M\"oller that produces larger rates,
a feature that is connected with the structure of the GT strength 
distribution in Fig. (\ref{bgt_v51}).

For  $^{54}$Fe (Fig. \ref{rate_fe54}) $Q_{EC}$ is again very small allowing
most of the GT strength to be probed more easily. At low densities, 
the agreement of our QRPA calculations with experiment is better than 
for other models.
It is also interesting to observe how the rates from QRPA-M\"oller are
much lower than other models at low $\rho$ and $T$, but are larger at
the largest density. This is a consequence of the GT distribution in
Fig. \ref{bgt_fe54} that shows a huge strength at 4 MeV and practically 
nothing else. Then, at low  $\rho$ and $T$ there are no electrons 
available to be captured to that energy. On the contrary at high densities,
when the Fermi energy is larger, strong EC to that state is possible
increasing dramatically the rate. 

In the case of $^{55}$Mn (Fig. \ref{rate_mn55}) the most remarkable
aspect to mention is the similarity between the model predictions,
as well as the discrepancy with experiment at low  $\rho$ and $T$.
This is caused by the larger experimental GT strength at low energies.
At higher $\rho$ when the effect of the low-lying states is not so
important the agreement with the experiment improves.
The situation is very similar in  $^{56}$Fe (Fig. \ref{rate_fe56}),
where the experimental GT strength at low excitation energy is
larger than any model prediction. The rates from QRPA-M\"oller 
are particularly low as a consequence of the lack of GT strength 
below 2.5 MeV (see Fig. \ref{bgt_fe56}).

In  $^{58}$Ni (Fig. \ref{rate_ni58}) with a very small $Q_{EC}$ we can
see that at low  $\rho$ the rates are sensitive to the strength of the 
low-lying excitations and thus the rates from QRPA-M\"oller are the 
smallest, while at the high $\rho$ the rates are sensitive to the 
total strength and then QRPA-M\"oller is the largest in accordance 
with the GT strengths in Fig. \ref{bgt_ni58}.
The situation in $^{59}$Co (Fig. \ref{rate_co59}) is similar to
the previous case, but in this case our QRPA rates are the 
lowest at low $\rho$ because the GT strengths are the smallest 
at low energies.

In the cases of $^{64}$Ni and $^{62}$Ni, and to a less extent $^{60}$Ni,
$Q_{EC}$ has large negative values and therefore the rates are mainly 
sensitive to the strength of the low-lying states. In the case of 
$^{60}$Ni (Fig. \ref{rate_ni60}) the rates from the different models 
at low $\rho$ vary from the larger values of the experimental rates up 
to the lower values of the QRPA-M\"oller calculations. This is in 
agreement with the GT strength at low energies, where the measured 
strength is larger than any model. The GT strength from QRPA-M\"oller 
is practically inexistent up to 2 MeV. 
At higher $\rho$ one finds the rates ordered from the 
higher QRPA to the lower SM, passing through the experimental rates.
In this case this ordering follows the total GT strength in Fig.
\ref{bgt_ni60} as one could expect from the fact that at high
$\rho$, the Fermi energy is high enough to overcome the range of
excitation energies where GT strength is found.
Similar arguments apply to $^{62}$Ni and $^{64}$Ni.
Finally, the rates in Fig. \ref{rate_zn64} for $^{64}$Zn with a very small 
$Q_{EC}$ energy, are sensitive to high excitation energies that can
contribute especially at high $\rho$ values.

\begin{figure}
\includegraphics[width=0.40\textwidth]{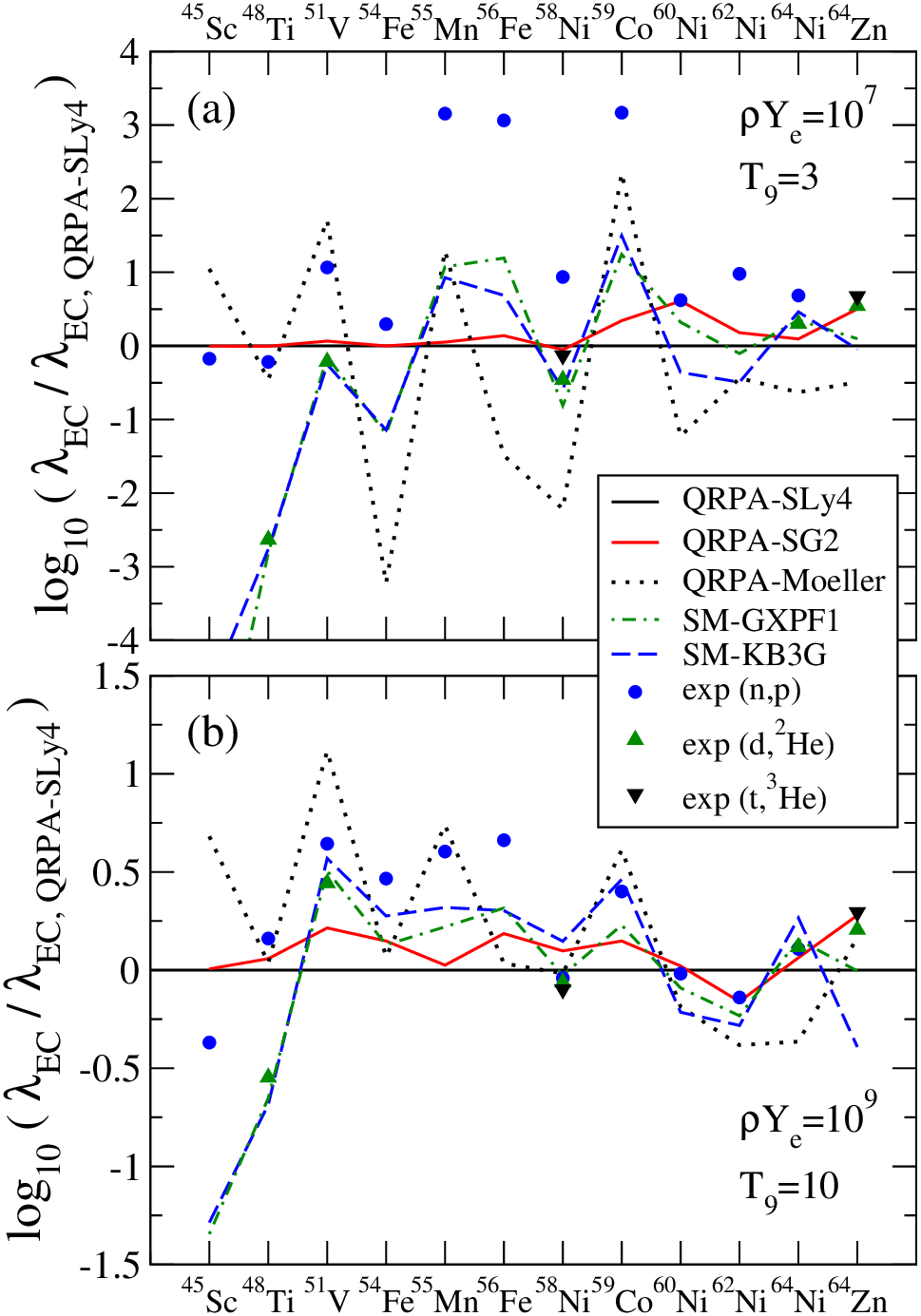}
\caption{(Color online) Ratios of various EC rates calculated from theoretical
and experimental GT strength distributions with respect to the EC rates
calculated in QRPA with the Skyrme force SLy4. The EC rates correspond to 
the stellar conditions (a) $\rho Y_e=10^7$ mol/cm$^3$, $T_9=3$ GK, and
(b) $\rho Y_e=10^9$ mol/cm$^3$, $T_9=10$ GK.}
\label{ratios}
\end{figure}

To quantify the measure of the quality of the various calculations for the 
EC rates, the ratios of the EC rates are compared in Fig. \ref{ratios} for 
the {\it pf}-shell nuclei studied in this work and for two stellar conditions
corresponding to $\rho Y_e=10^7$ mol/cm$^3$, $T_9=3$ GK in Fig. \ref{ratios}(a)
and to $\rho Y_e=10^9$ mol/cm$^3$, $T_9=10$ GK in Fig. \ref{ratios}(b).
Because we do not have experimental GT strength distributions from high
resolution charge-exchange reactions, $(d,^2{\rm He})$ or $(t,^3{\rm He})$,
for all of these nuclei, we show in this comparison the relative EC rates
using the QRPA-SLy4 results as a reference, 
$\log_{10}(\lambda_{\rm EC}/\lambda_{\rm EC, QRPA-SLy4})$.
As it was noticed in Ref. \cite{cole-12}, the results in Fig. \ref{ratios}
confirm that the SM calculations reproduce fairly well the rates from the high 
resolution data, while our QRPA results improve significantly the results from 
QRPA-M\"oller, except in the case of $^{48}$Ti. The comparison of the calculations 
with the rates obtained from $(n,p)$ data shows a tendency to underestimate them 
in most cases. It is also worth mentioning the better agreement with the data 
with the stellar conditions given in Fig. \ref{ratios}(b), where the EC rates 
are less sensitive to the details of the GT strength distributions.

The quality of the calculations can be further quantified with the help of
the average ratios of all nuclei defined by

\begin{equation}
\overline{\Lambda_{\rm EC}}=\frac{1}{N}\sum_i^N \log_{10} \left[ \lambda_i(\rm th) /
\lambda_i (\rm exp) \right]\, ,
\label{average1}
\end{equation}
or with the average of the absolute ratios that avoid positive and negative 
cancellations and gives us the average order of magnitude of these ratios,

\begin{equation}
\overline{ \left| \Lambda_{\rm EC}\right| } =\frac{1}{N}\sum_i^N \left| \log_{10} \left[
\lambda_i(\rm th) /\lambda_i (\rm exp) \right] \right| \, .
\label{average2}
\end{equation}

These average results are given in Tables III and IV for the two stellar 
conditions discussed above. In these tables one can see the ratios of various 
theoretical EC rates with respect to the experimental rates for {\it pf}-shell 
nuclei. We consider two sets of experimental rates, in the first row of
each model the ratios are calculated with respect to the rates obtained from
the GT strength distributions extracted from $(n,p)$ charge-exchange reactions.
In the second row the ratios are calculated with respect to the high resolution
charge exchange reactions (h.r.). 
The last two columns correspond to the average of the ratios and to the average
of the absolute ratios, as defined in Eqs. (\ref{average1}) and (\ref{average2}), 
respectively. In the case of $(n,p)$ data the average involves 11 nuclei, whereas
in the case (h.r.) there are 5 nuclei involved.

Table III shows that for these stellar conditions, the different theoretical 
models produce comparable $\overline{ |\Lambda_{\rm EC}|}$ average ratios in 
the case of $(n.p)$ data, with 
a little bit better agreement for the QRPA calculations with Skyrme forces. 
On the other hand, the best agreement in the case of high resolution data is 
obtained with the SM calculations, although the agreement in the QRPA-Skyrme 
cases is also quite good. In the stellar conditions of Table IV the quality
of the calculations improve and the agreement with experiment is much better. 
The average of the absolute ratios of the different models with respect to 
the $(n,p)$ data are comparable. On the other hand, the agreement with the high 
resolution data is especially good for the SM with the GXPF1 force. The agreement 
obtained from SM with KB3G and from QRPA with SLy4 and SG2 Skyrme forces is 
comparable, while the QRPA-M\"oller results are somewhat worse.

\begin{figure}
\includegraphics[width=0.40\textwidth]{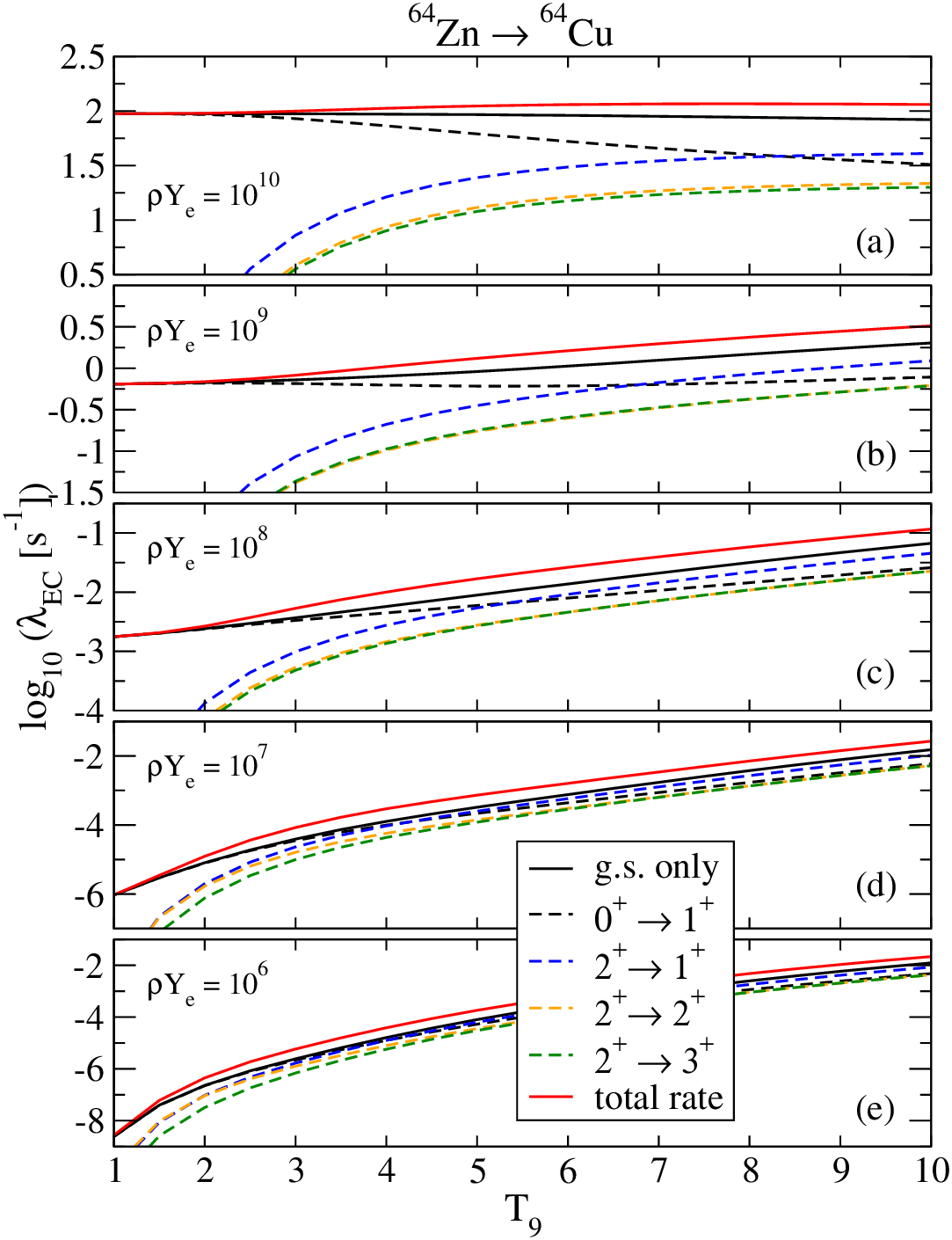}
\caption{(Color online)  Electron-capture rates in $^{64}$Zn
calculated from QRPA-SLy4 at various densities and temperatures.
The total rates are decomposed into their contributions from
decays of the ground state $0^+$ and the $2^+$ excited state.}
\label{misce_zn64}
\end{figure}

As it has been mentioned, ECs from excited states in the parent nuclei have
not been considered in this work on the basis of the typical excitation
energies of the lowest $2^+$ excited states that are in most cases larger 
than 1 MeV and the $T$-range considered that is not sufficiently high to 
populate those states significantly.

Nevertheless, in order to check to what extent these contributions
are small, we have calculated the EC rates originated by transitions 
from the $2^+$ excited state in the case of $^{64}$Zn, where 
$E_{2^+}= 0.992$ MeV \cite{ensdf} is one of the lowest excitation energy
in our set of nuclei. We compare in Fig. \ref{misce_zn64} 
these EC rates with those from the ground state in the case of QRPA-SLy4.
In this figure we can see the various contributions to the EC rates for 
several densities as a function of $T$ coming from the transitions
from the ground state $0^+_{\rm gs}\rightarrow 1^+$ and from
the transitions from the excited state $2^+ \rightarrow 1^+,2^+,3^+$ in the
parent nucleus, as well as the sum of these contributions (labeled as 
total rate). The figure also shows for comparison the EC rates calculated 
from the ground state alone with a probability of population set to one
(g.s. only), which correspond to the QRPA-SLy4 results in Fig. \ref{rate_zn64}.
The results show that EC from the ground state is the dominant contribution
except at high $T$, where contributions from excited states are comparable.
We also observe that the EC from only the ground state is always larger
than the contribution $0^+_{\rm gs}\rightarrow 1^+$ to the total rate because
of the depopulation of the ground state when the $2^+$ state becomes populated.


\begin{table}[ht]
\begin{center}
\caption{ Electron chemical potentials $\mu_e$ (MeV) for selected values 
of densities $\rho Y_e$ (mol/cm$^3$) and temperatures $T_9$ (GK).}
\vskip 0.5cm
\begin{tabular}{lccccc}
\hline \hline \\
$\rho Y_e$ & \multicolumn{5}{c}{$T_9$} \\
\cline{2-6} \\
         & 1   &  3    &   5  &  7   &  10 \\
\hline \\
$10^6$    &  0.672 & 0.299 &  0.091 &  0.042 &  0.020 \\
$10^8$    &  2.437 & 2.355 &  2.192 &  1.952 &  1.493 \\
$10^{10}$ & 11.116 & 11.098 & 11.063 & 11.011 & 10.898 \\
\hline \hline
\end{tabular}
\end{center}
\label{tableef}
\end{table}


\begin{table*}[ht]
\begin{center}
\caption{Ratios of various theoretical EC rates with respect to the 
experimental rates, $\log_{10}(\lambda_{\rm EC, th}/\lambda_{\rm EC, exp})$, 
for {\it pf}-shell nuclei. The experimental EC rates of reference correspond to 
the GT strength distributions extracted from $(n,p)$ or high resolution
charge exchange reactions (h.r.). The stellar density and temperature
conditions are $\rho Y_e=10^7$ mol/cm$^3$, $T_9=3$ GK. The last two 
columns correspond to the average of the ratios $\overline{ \Lambda_{\rm EC}}$
and to the average of the absolute ratios 
$\overline{ \left| \Lambda_{\rm EC} \right| }$, as defined in 
Eqs. (\ref{average1}) and (\ref{average2}), respectively. 
}
\vskip 0.5cm
\begin{tabular}{llccccccccccccccc}
\hline \hline \\
& &  $^{45}$Sc &  $^{48}$Ti & $^{51}$V  & $^{54}$Fe & $^{55}$Mn & $^{56}$Fe &
  $^{58}$Ni &  $^{59}$Co & $^{60}$Ni & $^{62}$Ni & $^{64}$Ni & $^{64}$Zn & \qquad & 
$\overline{ \Lambda_{\rm EC}}$ & $\overline{ \left| \Lambda_{\rm EC} \right| }$ \\ \\
SM-KB3G & $(n,p)$ & -4.518 & -2.544 & -1.320 & -1.439 & -2.228 & -2.376 & -1.526 &
 -1.671 & -0.980 & -1.470 & -0.223 &  && -1.845 & 1.845 \\
& (h.r.) &   & -0.130 & -0.040 &  &  &  & -0.460 &
  &  &  & 0.159 & -0.589 && -0.212 & 0.276 \\
SM-GXPF1 & $(n,p)$ & -6.357 & -2.596 & -1.217 & -1.490 & -2.081 & -1.869 & -1.746 &
 -1.924 & -0.298 & -1.082 & -0.334 &  && -1.909 & 1.909 \\
& (h.r.) &   & -0.182 & 0.063 &  &  &  & -0.680 &
  &  &  & 0.048 & -0.446 && -0.239 & 0.284 \\
QRPA-M\"oller & $(n,p)$ & 1.219 & -0.240 & 0.642 & -3.518 & -1.864 & -4.551 & -3.151 &
 -0.820 & -1.849 & -1.417 & -1.316 &  && -1.533 & 1.872 \\
& (h.r.) &   & 2.174 & 1.922 &  &  &  & -2.084 &
  &  &  & -0.934 & -1.034 && 0.009 & 1.630 \\
QRPA-SLy4 & $(n,p)$ & 0.175 & 0.217 & -1.066 & -0.296 & -3.155 & -3.063 & -0.936 &
 -3.169 & -0.620 & -0.979 & -0.685 &  && -1.234 & 1.306 \\
& (h.r.) &   & 2.631 & 0.214 &  &  &  & 0.131 &
  &  &  & -0.303 & -0.543 && 0.426 & 0.764 \\
QRPA-SG2 & $(n,p)$ & 0.172 & 0.211 & -1.001 & -0.296 & -3.103 & -2.924 & -0.988 &
 -2.826 & -0.010 & -0.799 & -0.590 &  && -1.105 & 1.175 \\
& (h.r.) &   & 2.625 & 0.279 &  &  &  & 0.078 &
  &  &  & -0.208 & -0.041 && 0.547 & 0.646 \\
\hline \hline
\end{tabular}
\end{center}
\label{tablecase1}
\end{table*}


\begin{table*}[ht]
\begin{center}
\caption{Same as in Table III, but for the stellar conditions
 $\rho Y_e=10^9$ mol/cm$^3$, $T_9=10$ GK. 
}
\vskip 0.5cm
\begin{tabular}{llccccccccccccccc}
\hline \hline \\
 &&  $^{45}$Sc &  $^{48}$Ti & $^{51}$V  & $^{54}$Fe & $^{55}$Mn & $^{56}$Fe &
  $^{58}$Ni &  $^{59}$Co & $^{60}$Ni & $^{62}$Ni & $^{64}$Ni & $^{64}$Zn & \qquad &
$\overline{ \Lambda_{\rm EC}}$ & $\overline{ \left| \Lambda_{\rm EC} \right| }$ \\ \\
SM-KB3G & $(n,p)$ & -0.917 & -0.846 & -0.075 & -0.190 & -0.286 & -0.359 & 0.186 &
 0.061 & -0.197 & -0.142 & 0.158 &  && -0.237 & 0.311 \\
& (h.r.) &   & -0.138 & 0.126 &  &  &  & 0.245 &
  &  &  & 0.147 & -0.597 && -0.043 & 0.251 \\
SM-GXPF1 & $(n,p)$ & -0.976 & -0.813 & -0.139 & -0.337 & -0.385 & -0.346 & 0.014 &
 -0.172 & -0.073 & -0.093 & 0.047 &  && -0.298 & 0.309 \\
& (h.r.) &   & -0.105 & 0.062 &  &  &  & 0.072 &
  &  &  & 0.036 & -0.211 && -0.029 & 0.097 \\
QRPA-M\"oller & $(n,p)$ & 1.050 & -0.125 & 0.471 & -0.399 & 0.134 & -0.628 & 0.023 &
 0.211 & -0.169 & -0.242 & -0.472 &  && -0.013 & 0.357 \\
& (h.r.) &   & 0.583 & 0.672 &  &  &  & 0.082 &
  &  &  & -0.483 & -0.039 && 0.163 & 0.372 \\
QRPA-SLy4 & $(n,p)$ & 0.369 & -0.161 & -0.644 & -0.466 & -0.604 & -0.662 & 0.040 &
 -0.401 & 0.018 & 0.140 & -0.108 &  && -0.226 & 0.328 \\
& (h.r.) &   & 0.547 & -0.443 &  &  &  & 0.099 &
  &  &  & -0.118 & -0.206 && -0.024 & 0.283 \\
QRPA-SG2 & $(n,p)$ & 0.374 & -0.103 & -0.430 & -0.318 & -0.580 & -0.476 & 0.138 &
 -0.253 & 0.038 & -0.020 & -0.045 &  && -0.152 & 0.252 \\
& (h.r.) &   & 0.605 & -0.228 &  &  &  & 0.197 &
  &  &  & -0.055 & 0.073 && 0.118 & 0.232 \\
\hline \hline
\end{tabular}
\end{center}
\label{tablecase2}
\end{table*}

\section{Conclusions and final remarks}
\label{conclusions}

In this work we have evaluated continuum electron-capture rates at different 
density and temperature conditions holding in stellar scenarios. This study 
is performed on a set of {\it pf}-shell nuclei representative of the 
constituents in presupernova formations (i.e., $^{45}$Sc, $^{48}$Ti, $^{51}$V, 
$^{54}$Fe, $^{55}$Mn, $^{56}$Fe, $^{58}$Ni, $^{59}$Co, $^{60}$Ni, $^{62}$Ni, 
$^{64}$Ni, and $^{64}$Zn). The nuclear structure involved in the calculation 
of the energy distribution of the Gamow-Teller strength is described within 
a selfconsistent deformed HF+BCS+QRPA formalism with density-dependent 
effective Skyrme interactions and spin-isospin residual interactions.

We find that the present QRPA calculation is able to reproduce the main 
features of the GT distributions extracted in these nuclei from
charge-exchange reactions. Comparison of our results for the EC rates with 
SM calculations and other QRPA results analyzed in Ref. \cite{cole-12} shows 
that, in general, the agreement with experiment is best in the case of SM 
calculations, but the present QRPA calculations with Skyrme forces 
are of similar quality in most cases, clearly improving the agreement with 
experiment with respect to QRPA-M\"oller.
Thus, the results in this work provide additional information on the 
performance of QRPA-based models and refine the conclusions in 
Ref. \cite{cole-12}, where it was found that QRPA calculations based on 
the model developed in Ref. \cite{krumlinde-84} produce systematically 
much larger deviations from the data than SM calculations. This work 
demonstrates that this is not the case for all QRPA type calculations.

We have studied the sensitivity of the EC rates to both $\rho$ and $T$.
At low $\rho$ (low Fermi energies) and low $T$ (sharp shape of the energy 
distribution of the electrons), the rates are very sensitive to details 
of the GT strength of the low-lying excitations and therefore to model 
calculations. On the other hand, when the $\rho$ and $T$ are high enough, 
the  EC rates are sensitive to all the spectrum. Then, the whole description 
of the GT strength distribution is more important than a detailed description
of the low-lying spectrum. Since QRPA reproduces reasonably well the global 
behavior of the GT strength distributions, it is expected to be a good 
approach, especially for high $\rho$ and $T$ conditions.

$\beta ^+$ decays and EC from thermally populated excited states in the 
parent nuclei are expected to occur at high $T$. Contributions to the 
EC rates from the low-lying excited state in $^{64}$Zn have been evaluated
and compared with the contributions from the ground state in the range
of $\rho$ and $T$ considered in this work. A systematic study of these
contributions will be very important for more extreme stellar conditions,
but in this work they can be safely neglected. Moreover, the main
purpose of this work is to compare QRPA results with benchmark SM 
calculations and to EC rates extracted from the measured GT strength
distributions in the laboratory that do not contain those type of
contributions.


\acknowledgments

This work was supported in part by MINECO (Spain) under Research Grant
No.~FIS2011--23565 and by Consolider-Ingenio 2010 Programs CPAN 
CSD2007-00042.


\newpage

\begin{thebibliography}{00}

\bibitem{b2fh} E. M. Burbidge, G. R. Burbidge, W. A. Fowler, and F. Hoyle,
Rev. Mod. Phys. {\bf 29}, 547 (1957); 
G. Wallerstein  {\it et al.}, Rev. Mod. Phys. {\bf 69}, 995 (1997). 
\bibitem{ffn} G. M. Fuller, W. A. Fowler, and M. J. Newman,  
Ap. J. Suppl. {\bf 42}, 447 (1980); 
Ap. J. {\bf 252}, 715 (1982);  
Ap. J. Suppl. {\bf 48}, 279 (1982); 
Ap. J. {\bf 293}, 1 (1985).
\bibitem{aufderheide-94} M. B. Aufderheide, I. Fushiki, S. E. Woosley, and 
D. H. Hartmann, Ap. J. Suppl. {\bf 91}, 389 (1994);
A. Heger, S. E. Woosley, G. Mart\'{\i}nez-Pinedo, and  K. Langanke,
Ap. J. {\bf 560}, 307 (2001).  
\bibitem{langanke-03} K. Langanke and G. Mart\'{\i}nez-Pinedo,
Rev. Mod. Phys.  {\bf 75}, 819 (2003).
\bibitem{alford-91} W. P. Alford {\it et al.}, Nucl. Phys. {\bf A531}, 97 (1991).
\bibitem{yako-09} K. Yako {\it et al.}, Phys. Rev. Lett. {\bf 103}, 012503 (2009). 
\bibitem{rakers-04} S. Rakers {\it et al.}, Phys. Rev. C {\bf 70}, 054302 (2004).
\bibitem{alford-93} W.P. Alford {\it et al.}, Phys. Rev. C {\bf 48}, 2818 (1993).
\bibitem{baumer-03} C. B\"aumer {\it et al.}, Phys. Rev. C {\bf 68}, 031303(R) (2003).
\bibitem{ronnqvist-93} T. R\"onnqvist {\it et al.}, Nucl. Phys. {\bf A563}, 225 (1993).
\bibitem{vetterli-87} M. C. Vetterli {\it et al.}, Phys. Rev. Lett. {\bf 59}, 
439 (1987); Phys. Rev. C {\bf 40} 559 (1989).
\bibitem{elkateb-94} S. El-Kateb {\it et al.},  Phys. Rev. C {\bf 49}, 3128 (1994).
\bibitem{hagemann-04} M. Hagemann  {\it et al.}, Phys. Rev. C {\bf 71}, 014606 (2005);
Phys. Lett. {\bf B 579}, 251 (2004).
\bibitem{cole-06} A. L. Cole  {\it et al.}, Phys. Rev. C {\bf 74}, 034333 (2006).
\bibitem{williams-95} A. L. Williams {\it et al.},  Phys. Rev. C {\bf 51}, 1144 (1995).
\bibitem{anantaraman-08} N. Anantaraman  {\it et al.}, Phys. Rev. C {\bf 78}, 
065803 (2008).
\bibitem{popescu-07} L. Popescu {\it et al.}, Phys. Rev. C {\bf 75}, 054312 (2007).
\bibitem{grewe-08} E.-W. Grewe  {\it et al.}, Phys. Rev. C {\bf 77}, 064303 (2008).
\bibitem{hitt-09} G. W. Hitt  {\it et al.}, Phys. Rev. C {\bf 80}, 014313 (2009).
\bibitem{fujita-11} Y. Fujita, B. Rubio, and W. Gelletly, Prog. Part. Nucl. 
Phys. {\bf 66}, 549 (2011).
\bibitem{osterfeld-92} F. Osterfeld, Rev. Mod. Phys. {\bf 64}, 491 (1992).
\bibitem{koonin-97} S. E. Koonin, D. J. Dean, and K. Langanke, 
Phys. Rep. {\bf 278}, 1 (1997).
\bibitem{langanke-00} K. Langanke and G. Mart\'{\i}nez-Pinedo,
Nucl. Phys. {\bf A673}, 481 (2000).
\bibitem{langanke-01}  K. Langanke and G. Mart\'{\i}nez-Pinedo,
At. Data Nucl. Data Tables {\bf 79}, 1 (2001).
\bibitem{suzuki-09} T. Suzuki  {\it et al.}, Phys. Rev. C {\bf 79}, 061603(R) (2009); 
Phys. Rev. C {\bf 83}, 044619 (2011).
\bibitem{moeller-97} P. M\"oller, J. R. Nix, and K. -L. Kratz,
At. Data Nucl. Data Tables {\bf 66}, 131 (1997).
\bibitem{nabi-99} J.-U. Nabi and H. V. Klapdor-Kleingrothaus, 
At. Data Nucl. Data Tables {\bf 71}, 149 (1999); {\bf 88}, 237 (2004).
\bibitem{nabi-09} J.-U. Nabi, Eur. J. A. {\bf 40}, 223 (2009);
Astrophys. Space Sci. {\bf 331}, 537 (2010); 
Eur. J. A. {\bf 48}, 84 (2012).
\bibitem{paar-04} N. Paar, T. Nik\v{s}i\'c, D. Vretenar, and P. Ring,  Phys. 
Rev. C {\bf 69}, 054303 (2004); 
N. Paar, G. Col\`o, E. Khan, and D. Vretenar, Phys. Rev. C {\bf 80}, 055801 (2009).
\bibitem{sarri-09-11} P. Sarriguren,  Phys. Lett. {\bf B 680}, 438 (2009);
Phys. Rev. C {\bf 83}, 025801 (2011).
\bibitem{fantina-12} A. F. Fantina, E. Khan, G. Col\`o, N. Paar, and D. Vretenar,
Phys. Rev. C {\bf 86}, 035805 (2012).
\bibitem{cole-12} A. L. Cole {\it et al.}, Phys. Rev. C {\bf 86}, 015809 (2012).
\bibitem{poves-01} A. Poves, J. S\'anchez-Solano, E. Caurier, and F. Nowacki, 
Nucl. Phys. {\bf A694}, 157 (2001).
\bibitem{honma-02} M. Honma, T. Otsuka, B. A. Brown, and T. Mizusaki, Phys. Rev.
C {\bf 65}, 061301(R) (2002); {\bf 69}, 034335 (2004).
\bibitem{krumlinde-84} J. Krumlinde and P. M\"oller, Nucl. Phys. {\bf A417}, 419 (1984); 
P. M\"oller and J. Randrup, Nucl. Phys. {\bf A514}, 1 (1990).
\bibitem{moeller-95} P. M\"oller, J. R. Nix, W. D. Myers, and W. J. Swiatecki,
At. Data Nucl. Data Tables {\bf 59}, 185 (1995).
\bibitem{halbleib-67} J. A. Halbleib and R. A. Sorensen, 
Nucl. Phys. {\bf A98}, 542 (1967).
\bibitem{muto-89}  K. Muto, E. Bender, and H.V. Klapdor, 
Z. Phys. A {\bf 333}, 25 (1989); 
K. Muto, E. Bender, T. Oda, and H.V. Klapdor-Kleingrothaus, 
Z. Phys. A {\bf 341}, 407 (1992).
\bibitem{frisk-95} F. Frisk, I. Hamamoto, and X. Z. Zhang, 
Phys. Rev. C {\bf 52}, 2468 (1995). 
\bibitem{sarri-98} P. Sarriguren, E. Moya de Guerra, A. Escuderos, and A. C. Carrizo, 
Nucl. Phys. {\bf A635}, 55 (1998).
\bibitem{sarri-99}  P. Sarriguren, E. Moya de Guerra, and A. Escuderos, 
Nucl. Phys. {\bf A658}, 13 (1999).
\bibitem{sarri-01} P. Sarriguren, E. Moya de Guerra, and A. Escuderos,
Nucl. Phys. {\bf A691}, 631 (2001).
\bibitem{sarri-01-odd} P. Sarriguren, E. Moya de Guerra, and A. Escuderos,
Phys. Rev. C {\bf 64}, 064306 (2001).
\bibitem{borzov-06} I. N. Borzov, Nucl. Phys. {\bf A 777}, 645 (2006).
\bibitem{sarri-03} P. Sarriguren, E. Moya de Guerra, R. \'Alvarez-Rodr\'{\i}guez,
Nucl. Phys. {\bf A716}, 230 (2003).
\bibitem{poirier-04} E. Poirier {\it et al.},  Phys. Rev. C {\bf 69}, 034307 (2004).
\bibitem{nacher-04} E. N\'acher {\it et al.},  Phys. Rev. Lett. {\bf 92}, 232501 (2004).
\bibitem{sarri-05-wp} P. Sarriguren,  R. \'Alvarez-Rodr\'{\i}guez, and E. Moya de Guerra, 
Eur. Phys. J. A {\bf 24}, 193 (2005).
\bibitem{sarri-09-prc} P. Sarriguren,  Phys. Rev. C {\bf 79}, 044315 (2009).
\bibitem{sarri-10} P. Sarriguren and J. Pereira,  Phys. Rev. C {\bf 81}, 064314 (2010).
\bibitem{ensdf} Evaluated Nuclear Structure Data File (ENSDF), 
  http://www.nndc.bnl.gov/ensdf/.
\bibitem{sly4} E. Chabanat, P. Bonche, P. Haensel, J. Meyer, and R. Schaeffer, 
Nucl. Phys. {\bf A635}, 231 (1998).
\bibitem{giai} N. Van Giai and H. Sagawa, Phys. Lett. {\bf B 106}, 379 (1981).
\bibitem{vautherin}  D. Vautherin and D. M. Brink, Phys. Rev. C {\bf 5}, 626 (1972); 
D. Vautherin, Phys. Rev. C {\bf 7}, 296 (1973).
\bibitem{flocard-73} H. Flocard, P. Quentin, A. K. Kerman, and D. Vautherin,
Nucl. Phys. {\bf A203}, 433 (1973).
\bibitem{homma-96} H. Homma {\it et al.}, Phys. Rev. C {\bf 54}, 2972 (1996).
\bibitem{bm} A. Bohr and B. Mottelson, {\em Nuclear Structure}, (Benjamin, New York 1975).
\bibitem{radha-97} P.B. Radha, D.J. Dean, S.E. Koonin, K. Langanke, and P. Vogel, 
Phys. Rev. C {\bf 56}, 3079 (1997).
\bibitem{caurier-99} E. Caurier, K. Langanke, G. Mart\'{\i}nez-Pinedo, and F. Nowacki, 
Nucl. Phys. {\bf A653}, 439 (1999).
\bibitem{hix-03} W. R. Hix {\it et al.}, Phys. Rev. Lett. {\bf 91}, 201102 (2003).
\bibitem{iwamoto-99} K. Iwamoto {\it et al.}, Ap. J. Suppl. {\bf 125}, 439 (1999).

\end{thebibliography}
\end{document}